\documentclass{comnet}
\usepackage{float}
\usepackage{graphicx}
\usepackage{arydshln}
\usepackage{amsmath}
\usepackage{amssymb}
\usepackage{amsthm}
\usepackage{todonotes} 
\usepackage{multirow}
\usepackage{array}
\usepackage{url}
\usepackage{hyperref}
\usepackage{cleveref}

\usepackage{bm}

\newcounter{minicaptionctr}
\newcommand{\minicaption}[1]{%
\refstepcounter{minicaptionctr}{\footnotesize \alph{minicaptionctr}) #1}}

\usepackage{tikz}
\usetikzlibrary{arrows.meta,positioning}
\usetikzlibrary{backgrounds}

\theoremstyle{plain}
\theoremstyle{definition}

\newcommand{\rightarrowsmall}{{_{^{^{\,\rightarrow}}}}}
\newcommand{\ord}[1]{^{(#1)}}
\newcommand{\curly}[1]{\mathcal{#1}}

\begin{document}
\title{Performance of Higher-Order Networks in Reconstructing Sequential Paths: from Micro to Macro Scale}


\author{
\name{Kevin Teo}
\address{Network Science Institute, Northeastern University London, London E1W 1LP, United
Kingdom}
\email{kt3512phd@students.nulondon.ac.uk}
\name{Naomi Arnold}
\address{Network Science Institute, Northeastern University London, London E1W 1LP, United
Kingdom}
\email{naomi.arnold@nulondon.ac.uk}
\name{Andrew Hone}
\address{School of Mathematics, Statistics \& Actuarial Science, University of Kent, Canterbury CT2 7NF, United Kingdom}
\email{A.N.W.Hone@kent.ac.uk}
\and 
\name{Istv\'an Zolt\'an Kiss}
\address{Network Science Institute, Northeastern University London, London E1W 1LP, United
Kingdom}
\email{istvan.kiss@nulondon.ac.uk}}
\date{July 2024}

\maketitle






\begin{abstract}
{Activities such as the movement of passengers and goods, the transfer of physical or digital assets, web navigation and even successive passes in football, result in timestamped paths through a physical or virtual network. The need to analyse such paths has produced a new modelling paradigm in the form of higher-order networks which are able to capture temporal and topological characteristics of sequential data. This has been complemented by sequence mining approaches, a key example being sequential motifs measuring the prevalence of recurrent subsequences. Previous work on higher-order networks has focused on how to identify the optimal order for a path dataset, where the order can be thought of as the number of steps of memory encoded in the model. In this paper, we build on these approaches to consider which orders are necessary to reproduce different path characteristics, from path lengths to counts of sequential motifs, viewing paths generated from different higher-order models as null models which capture features of the data up to a certain order, and randomise otherwise. Furthermore, we provide an important extension to motif counting, whereby cases with self-loops, starting nodes, and ending nodes of paths are taken into consideration. Conducting a thorough analysis using path lengths and sequential motifs on a diverse range of path datasets, we show that our approach can shed light on precisely where models of different order overperform or underperform, and what this may imply about the original path data.}
{Higher-Order Networks, Sequential Motifs, }\\ 
\end{abstract}

\section{Introduction} \label{Introduction}

In recent decades, technological advancements have pushed our world into an increasingly information-rich environment. Data collection methods have been rapidly developed and adopted into our everyday lives. In particular, the proliferation of the internet and mobile devices has enabled the collection of more accurate temporal data in a wide variety of sectors ranging from shipping and finance to sports and travel. With the improvement of time-resolved data into shorter and more precise timescales, the study of temporal correlations has become a topic of great interest and importance with regard to the modelling and prediction of dynamical complex systems \cite{lambiotte2019networks, chawla2016representing, benson2021higher, rosvall2014memory}.  


Specifically, there has been an immense growth in the availability and use of sequential data, where the sequential order is imposed by time. Sequential data is often recorded as an ordered list of events,items, or locations, sometimes with associated timestamps. Popular approaches to studying such data include network theory, where the network encodes binary relations between the elements of the sequential data, allowing for the use of network-based measures and metrics for the analysis of such systems.  

Synthetic sequences can be simulated via the movement of agents on the network itself, mathematically known as a random walk. This has been applied to complex real-world multi-agent systems \cite{rosvall2014memory, peixoto2017modelling}, such as the movement of users between pages on a website \cite{larock2022sequential}. Classically or in the simplest approximation, the frequency of activity between two nodes is used as a proxy to determine the next destination of an agent. However, this may not always match reality, since real-world agents typically make decisions using both present and historical information to determine their next choice. This raises the question of what mathematical abstraction is appropriate to cater for history-dependent decisions or movement.  Indeed, it has been shown that the standard network approach to modelling such data often fails to capture the observed temporal dependencies and correlations \cite{scholtes2017network, chawla2016representing, edler2017mapping}. 



%
%

Variable/multi/higher-order networks were devised to capture both the underlying network topology and the path history from sequential data within a single underlying mathematical framework \cite{scholtes2017network, chawla2016representing, saebi2020efficient, tonon2023caspita, bick2023higher}, by utilising the mathematics of variable order Markov chains \cite{buhlmann1999variable} in a network context.  These models allow for the retention of the most recent node history of a particular length, prescribed by the order.  Prior work done by Scholtes et al. \cite{scholtes2017network, petrovic2022learning} with the \verb|PathPy| package focused on determining an optimal highest order, i.e. how much recent memory is required in a process in order to adequately model the system, balancing the explanatory power via the maximum likelihood versus the complexity of the model, harnessing the nestedness properties of their model to apply Wilks' theorem in a likelihood ratio test. Further work with Gote et al. via the Multi Order Generative (MOGen)\cite{gote2020predicting, gote2023predicting} model improved on \verb|PathPy| and also includes an order detection method based on the Akaike Information Criterion \cite{aic}. These models also include other information metrics to evaluate the model, such as the cross-entropy loss from next-element prediction. Other models such as BuildHON\cite{chawla2016representing} and BuildHON+\cite{saebi2020efficient} approach generating higher order nodes if and only if the outgoing-edge distribution of the higher order node is significantly different from the lower order node, where the statistical difference is measured with the Kullback-Leibler divergence \cite{kullback1951information}. 


The detection of the optimal order via information theory criteria ignores the evaluation of how representative the model is at explaining other non-information theoretic features of the data. For example, we show that modelling data at an optimal order (detected by MOGen for example) does not guarantee that other features of the data, such as the total sequence length prediction, will be accurately represented by the model. Of particular interest for our investigation into categorical sequential data is the study of sequential motifs. These sequential motifs - henceforth usually referred to as simply ``motifs'' - can reveal path traversal patterns that are common among agents. For example, consider flight itineraries of passengers from all over the world. Regardless of their specific origin, many passengers will book a return flight from $origin \rightarrow destination \rightarrow origin$ \cite{larock2022sequential, edler2017mapping}. This represents a temporal dependency that favours returning to the state $1$ time-step before the current location.

The paper is structured as follows: In section \ref{Model}, we first introduce the concept of sequential motifs and their significance in the study of sequential data. We then cover variable-order Markov chains to define the higher-order network model, along with its corresponding likelihood function. The model additionally accounts for the start and end of sequences, thereby aiming to capture the dynamics induced by transitions from activity to inactivity of agents and processes on the network. This is followed by a discussion of several model selection techniques often utilised to identify an optimal order for the model. In section \ref{Results}, we present the details of several datasets to which the higher-order models are applied, in order to evaluate the performance of these models at various orders, including any optimal order chosen by other model selection techniques. We focus primarily on the sequence length distribution and motif frequency distribution, which are macro- and meso-scale features,  respectively. Algorithmically, we implement these models at the $1^{st}$, $2^{nd}$, $3^{rd}$ and $4^{th}$ order, and define appropriate metrics, such as the Kolmogorov-Smirnov test for comparing sequence length distributions, and Studentised Residuals for comparing motif frequencies. Finally, in section \ref{Conclusion}, we provide additional discussion of the work done, potential areas of applicability, and further research.

\section{Model and Methods} \label{Model}

We suppose that the given data consists of a multiset $\curly D$ of ordered sequences $S = (v_1 \rightarrowsmall \cdots \rightarrowsmall v_{|S|})$ where $|S|$ is the number of elements in the sequence, and each element $v_j\in\curly V$, where $\curly V$ is the set of nodes in a network. Such sequences are often referred to as \textit{walks}. A sequence with no repeated elements is also referred to as a \textit{path}. The length $l$ of a sequence $S$ is the number of steps or transitions in the sequence, so that $|S| = l+1$ and $S = (v_1 \rightarrowsmall \cdots \rightarrowsmall v_{l+1})$. 

A finite sequence $S'=(v_1'\rightarrowsmall\cdots \rightarrowsmall v_{l'+1}')$ of length 
$l'\leq l$ is a sub-sequence of $S$ if there exists an integer $i\geq 0$ such that $\forall j \in (1, \ldots, l'+1), v'_j = v_{i+j}$. For convenience, we refer to a sub-sequence of length $1$ as a `hop'. 

In addition to sequences, we can also construct the basic weighted directed network $\curly G = (\curly V, \curly E)$ with the node set $\curly V$ and edge set $\curly E \subseteq \curly V \times \curly V$, with the following properties:
\begin{itemize}
    \item $\forall v \in \curly V, \exists S \in \curly D$ such that $v \in S$, and
    \item $\forall (u \rightarrowsmall v) \in \curly E: \exists S \in \curly D$ such that $(u \rightarrowsmall v)$ is a sub-sequence in $S$;  and we can further assign a weight $W\in \mathbb{R}$ to each edge.  
\end{itemize}
In section~\ref{subsec:likelihood}, we assign the edge weights to represent observation frequencies and transition probabilities. Generally in such problems, we assume \textit{a priori} that the topology of the basic weighted directed network is a true feature of the system described by the data. As such, all edges that are \textit{not} present in this network are assumed to be impossible. Conversely, all other combinations of traversing the network using the edges present are assumed to be possible, even if they are not observed in the data. 

\subsection{Sequential Motifs}


In recent years there has been renewed interest in understanding how dynamical processes unfold on a network across various domains of applicability. In travel data, for example, people often return to their origin; for cargo and other goods transportation, paths are often dictated by supply and demand pressures. There is scope to investigate whether these processes are representative of the temporal correlations that emerge from a real-world agent: by observing the frequencies of repeating sub-graphs within the network, known as motifs \cite{milo2002network}, small-scale patterns and structures can be identified and analysed. Sequential motifs take the directedness of the network into account, increasing the specificity of these patterns at the expense of a larger parameter space. 

Sequential motifs - hereon referred to simply as motifs - are a way of identifying patterns within sequences that are commonly observed in  data \cite{paranjape2017motifs, kovanen2011temporal}. A motif $M$ of length $l$ may be described as a sequence of symbols $(m_1, m_2, \ldots, m_{l+1})$ with $m_i \in \curly{A}$ for some fixed symbol set $\curly A$. 
A sequence $S = (v_1 \rightarrowsmall\cdots \rightarrowsmall v_{l+1})$ is an instance of the motif $M = (m_1, \ldots, m_{l+1})$ if for that sequence $S$ 
there is a function $f: \curly V \rightarrow \curly A$ with $f(v_i) = m_i$ for  $1\leq i\leq l+1$, and $m_i=m_j\implies v_i=v_j$, so $f$ is injective. 
For example, we can consider sequences representing regional locations of ships:  
\begin{example}[2-hop motifs from shipping sequences]
\begin{equation*}
    \begin{aligned}
        \big(f(\text{Mediterranean}) \rightarrow f(\text{South East Asia}) \rightarrow f(\text{Mediterranean})\big) = (A,B,A) \\
        \big(f(\text{China}) \rightarrow f(\text{South East Asia}) \rightarrow f(\text{China})\big) = (A,B,A) \\
        \big(f(\text{Red Sea}) \rightarrow f(\text{Baltic}) \rightarrow f(\text{North Sea})\big) = (A,B,C)
    \end{aligned}
\end{equation*}
Above we have taken the symbol set to be the alphabet 
$\curly A = \{ A,B,C,\ldots\}$, and for each set $S$ the function $f$ is chosen to assign each new element $v_i$ that appears to the next letter in $\curly A$.
\end{example}
Note that we also consider sequences and networks with self-loops, i.e. where $A
\rightarrowsmall A$ transitions are allowed. While self-loops may not seem intuitive in the context of physical movement through a network, they may arise as a way of representing an agent remaining in a location for a time step, or through a merging together of nodes. For example, in the shipping dataset used later in our experiments, the nodes originally representing ports were merged into 26 regions. In that context, a self-loop represents a ship moving between two ports within the same region. An example in virtual networks where self-loops make sense is modelling sequences of tokens/letters~\cite{peixoto2017modelling} where self-loops are a double letter.

\begin{figure}
    \begin{minipage}[b]{0.22\textwidth}
        \centering
        \begin{tikzpicture}[
              mycircle/.style={
                 circle,
                 draw=black,
                 fill=gray,
                 fill opacity = 0.3,
                 text opacity=1,
                 inner sep=0pt,
                 minimum size=20pt,
                 font=\small},
              myarrow/.style={-Stealth},
              node distance=0.2cm and 0.5cm
              ]
              \node[mycircle] (c1) {$A$};
              \node[mycircle,right=of c1] (c2) {$B$};
              \node[mycircle,right=of c2] (c3) {$C$};
        
            \foreach \i/\j/\txt/\p in {
              c1/c2/1/below,
              c2/c3/2/below}
               \draw [myarrow] (\i) -- node[sloped,font=\small,\p] {\txt} (\j);
            \end{tikzpicture}\\
        \minicaption{$A \rightarrowsmall B \rightarrowsmall C$}
    \end{minipage}
    \begin{minipage}[b]{0.18\textwidth}
        \centering
        \begin{tikzpicture}[
              mycircle/.style={
                 circle,
                 draw=black,
                 fill=gray,
                 fill opacity = 0.3,
                 text opacity=1,
                 inner sep=0pt,
                 minimum size=20pt,
                 font=\small},
              myarrow/.style={-Stealth},
              node distance=0.6cm and 0.8cm
              ]
              \node[mycircle] (c1) {$A$};
              \node[mycircle,right=of c1] (c2) {$B$};
        
            \draw[->] (c1) to[bend left] (c2);
            \draw[->] (c2) to[bend left] (c1);
            \path[->] (c1) to[bend left] node[midway,above,inner sep=2pt] {1} (c2);
            \path[->] (c2) to[bend left] node[midway,below,inner sep=2pt] {2} (c1);
            \end{tikzpicture}\\
        \minicaption{$A \rightarrowsmall B \rightarrowsmall A$}
    \end{minipage}
    \begin{minipage}[b]{0.18\textwidth}
        \centering
        \begin{tikzpicture}[
              mycircle/.style={
                 circle,
                 draw=black,
                 fill=gray,
                 fill opacity = 0.3,
                 text opacity=1,
                 inner sep=0pt,
                 minimum size=20pt,
                 font=\small},
              myarrow/.style={-Stealth},
              node distance=0.6cm and 0.8cm
              ]
              \clip (-2.25,-0.5) rectangle (0.5, 1.5);
              \node[mycircle] (c2) {$B$};
              \node[mycircle, left=of c2] (c1) {$A$};
        
            \draw[->] (c1) to[loop] ();
            \draw[->] (c1) to (c2);
            \path[->] (c1) to[loop] node[midway,above,inner sep=2pt] {1} ();
            \path[->] (c1) to node[midway,above,inner sep=2pt] {2} (c2);
            \end{tikzpicture}
        \minicaption{$A \rightarrowsmall A \rightarrowsmall B$}
    \end{minipage}
    \begin{minipage}[b]{0.18\textwidth}
        \centering
        \begin{tikzpicture}[
              mycircle/.style={
                 circle,
                 draw=black,
                 fill=gray,
                 fill opacity = 0.3,
                 text opacity=1,
                 inner sep=0pt,
                 minimum size=20pt,
                 font=\small},
              myarrow/.style={-Stealth},
              node distance=0.6cm and 0.8cm
              ]
              \clip (-0.5,-0.5) rectangle (2.25, 1.5);
              \node[mycircle] (c1) {$A$};
              \node[mycircle,right=of c1] (c2) {$B$};
        
            \draw[->] (c2) to[loop] ();
            \draw[->] (c1) to (c2);
            \path[->] (c2) to[loop] node[midway,above,inner sep=2pt] {2} ();
            \path[->] (c1) to node[midway,above,inner sep=2pt] {1} (c2);
            \end{tikzpicture} \hspace{-5mm}
        \minicaption{$A \rightarrowsmall B \rightarrowsmall B$}
    \end{minipage}
    \begin{minipage}[b]{0.18\textwidth}
        \centering
        \begin{tikzpicture}[
              mycircle/.style={
                 circle,
                 draw=black,
                 fill=gray,
                 fill opacity = 0.3,
                 text opacity=1,
                 inner sep=0pt,
                 minimum size=20pt,
                 font=\small},
              myarrow/.style={-Stealth},
              node distance=0.6cm and 0.8cm
              ]
              \clip (-0.6,-0.5) rectangle (0.6, 1.5);

              \node[mycircle] (c1) {$A$};
        
            \draw[->] (c1) to[loop] ();
            \draw[->] (c1) to[loop] ();
            \path[->] (c1) to[loop] node[midway,above,inner sep=2pt] {1} ();
            \path[->] (c1) to[loop] node[midway,below,inner sep=2pt] {2} ();
            \end{tikzpicture}\\
        \minicaption{$A \rightarrowsmall A \rightarrowsmall A$}
    \end{minipage}
    \caption{All possible two-edge sequential motifs with self-loops allowed.}
    \label{fig:motif-types}
\end{figure}
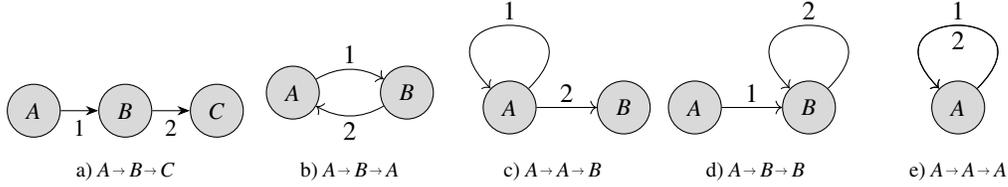

\begin{example}[1- and 2-hop sequential motifs]
    The only 1-hop motifs possible are all isomorphic to $(A,B)$ (an edge) or $(A,A)$ (a self-loop). There are five 2-hop motifs, all isomorphic to $(A,B,C)$, $(A,B,A),(A,B,B), (A,A,B)$ and $(A,A,A)$. These are shown in \cref{fig:motif-types}
\end{example}   

For a given length $l= |S|-1$, 
there is  a fixed number of motif permutations, denoted 
$a_{|S|}=a_{l+1}$ (so the above example shows 
$a_2=2$ and $a_3=5$). This corresponds to the number of ways to partition a set of $|S|$ labelled objects into unlabelled non-empty subsets, well known in combinatorics as the Bell numbers (or exponenential numbers) \cite{oeis}- see {\href{https://oeis.org/A000110}{\tt oeis.org/A000110}}. The Bell numbers are obtained from the recurrence relation 
\begin{equation*}
    a_{l+1} = \sum_{j=0}^{l}  \tbinom{l}{j}
    \,\, a_j, \quad  \text{with } a_0 = a_1 = 1.
\end{equation*}

\subsection{Variable Order Markov Chains}

At the beginning of this section, we defined variable length sequences, which may inherently contain higher order correlations. For example, if we only consider the frequency of movements between all possible node pairs and generate new sequences, this effectively ignores any potential sequential correlations. However, if at every node in the sequence, we concatenate it with the previous $k$ nodes, we can encode more of the recent history of the sequence. A priori, one would not know what the correlation length $k$ should be, e.g should we model paths as $(v_i \rightarrowsmall v_j) \rightarrow (v_j \rightarrowsmall v_k)$ or as $(v_i \rightarrowsmall v_j \rightarrowsmall v_k) \rightarrow (v_j \rightarrowsmall v_k \rightarrowsmall v_l)$? Furthermore, previous research has indicated an additional need to account for the variability in sequence length, to ensure that sequences shorter than the selected historical correlation length are meaningfully modelled. This problem has been studied in the literature through variable order Markov chains, which provides the mathematical foundation for the higher-order network representation. 

The basic model is the standard Markov chain \cite{grimmett2014probability}, which has been extensively used in sequence modelling as a next-element prediction task. It is formally defined by the transition probabilities 
\begin{equation}\label{smarkov}
    P(v_{i} | v_{1} \rightarrowsmall \cdots \rightarrowsmall v_{i-1}) = 
    P(v_{i} | v_{i-1}), 
\end{equation}
%
defined in terms of the elements in a sequence $S = (v_1 \rightarrowsmall \cdots \rightarrowsmall v_{l+1})$.
This is a `memoryless' Markov model, where only the current element (state) is known, and carries no memory of which  prior elements have passed. We can generalise this to higher-order Markov chains that predict the next element of the sequence based on the previous $k$ elements \cite{buhlmann1999variable, deshpande2004selective}, 
so that the transition probabilities satisfy the following:

\begin{equation}
\begin{aligned}
    P(v_i| v_1 \rightarrowsmall \cdots \rightarrowsmall v_{i-1}) =
    \begin{cases}
          P(v_i| v_{i-k} \rightarrowsmall \cdots \rightarrowsmall v_{i-1}) & \text{for } k \geq 1 ,\\
          P(v_i) & \text{for } k = 0.
    \end{cases}
\end{aligned}
\end{equation}
The integer $k$ is called the order: it corresponds to having a short-term  memory of length  $k$ influencing each transition probability. 
The standard (memoryless) Markov chain (\ref{smarkov}) has order $k=1$.  

Written explicitly, we can treat nodes in a $k$-length `memory' as a single $k$-vector node, that is
$$
\vec{v}\ord{k}_i = (v_{i-k+1} \rightarrowsmall \ldots \rightarrowsmall v_{i}), 
$$
and refer to this representation as a higher order node - in this case, an order $k$ node. A standard Markov chain  transition probability between two order $k$ nodes is then equivalent to a  higher-order Markov transition probability, of $k^{th}$ order:
\begin{equation}
    P\big(\vec{v}\ord{k}_i | \vec{v}\ord{k}_{i-1}\big) = P(v_i| v_{i-K} \rightarrowsmall \cdots \rightarrowsmall v_{i-1}).
    \label{higher-order-nodes}
\end{equation}

While transition probabilities allow us to construct a sequence node-by-node, they do not provide any additional insight into which element sequences begin from and when sequences end. This is crucial in constructing finite-length sequences, where how sequences start and end may greatly affect the properties of the sequence, such as the total length of the sequence, the number of unique nodes visited,~etc. A lack of information about how sequences start and end hinders the modelling of agents becoming active or inactive on the network in a realistic manner. We account for this by introducing an initial and a final state, denoted $*$ and 
$\dagger$ respectively, to account for the start and end of sequences. The initial transition probability is then given as $P(v_1|*)$. The final transition probability is given as $P(\dagger | v_{i-k+1} \rightarrowsmall \ldots \rightarrowsmall v_{i})$ for a $k^{th}$ order Markov chain. This can be represented in the sequence itself by appending $*$ and $\dagger$ before $v_1$ and after $v_{|S|}$ respectively: $S = (v_1 \rightarrowsmall \cdots \rightarrowsmall v_{|S|}) \rightarrow (* \rightarrowsmall v_1 \rightarrowsmall \cdots \rightarrowsmall v_{|S|} \rightarrowsmall \dagger)$. 
%
%

Combining the higher order Markov transition probabilities with the initial and final state transition probabilities, we can map any sequence $S = (v_1 \rightarrowsmall \cdots \rightarrowsmall v_{|S|})$ to a multi-layer higher order sequence $\tilde S$ with maximum order $k$, that is  

\begin{equation}
    * \rightarrow \overbrace{v_1}^{\text{order 1}} \rightarrow \overbrace{(v_1, v_2)}^{\text{order 2}} \rightarrow \cdots \rightarrow \overbrace{(v_1, \ldots, v_k) \rightarrow (v_2, \ldots, v_{k+1}) \rightarrow \cdots \rightarrow (v_{|S|-k + 1}, \ldots, v_{|S|})}^\text{{order k}} \rightarrow \dagger,
    \label{higher-order-sequence}
\end{equation}
using the higher order node representations described in equation (\ref{higher-order-nodes}), and from  $\curly D$ we can construct $\tilde{\curly D}$, the corresponding collection of higher order sequences. A convenient result of this mapping is that the length of $\tilde S$ remains the same at every order, and is equal to the original sequence length when ignoring $*$ and $\dagger$.

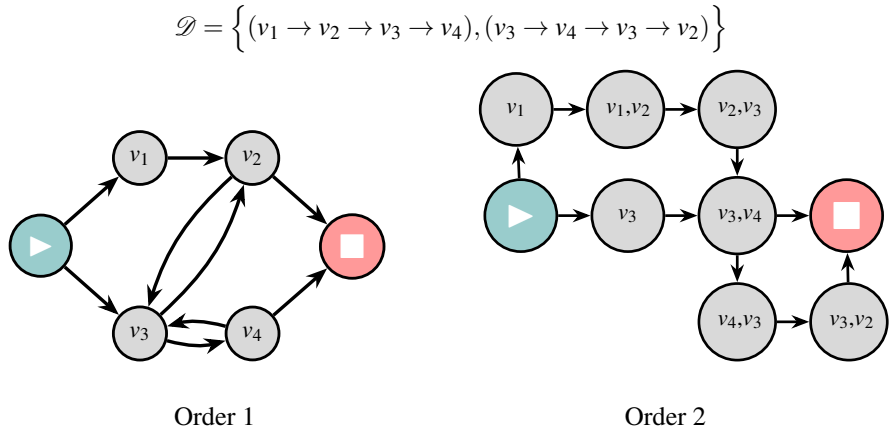
\begin{figure}[H]
    \centering
    $\curly D = \Big \{ (v_1 \rightarrow v_2 \rightarrow v_3 \rightarrow v_4 ), ( v_3 \rightarrow v_4 \rightarrow v_3 \rightarrow v_2) \Big \}$\\ \hfill \linebreak
    \resizebox{0.35 \linewidth}{!}{\begin{tikzpicture}[
      mycircle/.style={
         circle,
         draw=black,
         fill=gray,
         fill opacity = 0.3,
         text opacity=1,
         inner sep=5pt,
         minimum size=25pt,
         line width = 0.5mm, 
         font=\large},
      myarrow/.style={-Stealth},
      node distance=0.8cm and 1cm, line width=0.6mm
      ]
      \node[mycircle, fill = teal, font = \LARGE, fill opacity = 0.4] (a1) {$\textcolor{white}{\blacktriangleright}$};
      \node[mycircle, above right=of a1] (c2) {$v_1$};
      \node[mycircle, right=of c2] (c4) {$v_2$};
      \node[mycircle, below right=of a1] (d1) {$v_3$};
      \node[mycircle, right=of d1] (d2) {$v_4$};
      \node[mycircle, fill = red, font = \LARGE, fill opacity = 0.4, below right=of c4] (ax){$\textcolor{white}{\blacksquare}$};

     \foreach \i/\j/\txt/\p in {
      a1/d1/0.0/above,
      a1/c2/0.0/above,
      c2/c4/0.0/above,
      d2/ax/0.0/above,
      c4/ax/0.0/above}
       \draw [draw = black, myarrow] (\i) -- node[sloped,font=\small,\p] { } (\j);

       \path[draw = black, myarrow, every node/.style={font=\sffamily\small}]
    (d1) edge[bend right=15] node [left] {} (d2)
    (d2) edge[bend right=15] node [left] {} (d1)
    (c4) edge[bend right=15] node [left] {} (d1)
    (d1) edge[bend right=15] node [left] {} (c4);
\end{tikzpicture}} \hspace{10mm}
    \resizebox{0.38 \linewidth}{!}{\begin{tikzpicture}[
      mycircle/.style={
         circle,
         draw=black,
         fill=gray,
         fill opacity = 0.3,
         text opacity=1,
         inner sep=5pt,
         minimum size=38pt,
         line width = 0.5mm, 
         font=\large},
      myarrow/.style={-Stealth},
      node distance=0.5cm and 0.6cm, line width=0.5mm
      ]
      \node[mycircle, fill = teal, font = \huge, fill opacity = 0.4] (a1) {$\textcolor{white}{\blacktriangleright}$};
      \node[mycircle, right=of a1] (v3) {$v_3$};
      \node[mycircle, right=of v3] (v3v4) {{ $v_3$},$v_4$};
      \node[mycircle, above =of v3v4] (v2v3) {{ $v_2$},$v_3$};
      \node[mycircle, left=of v2v3] (v1v2){{ $v_1$},$v_2$};
      \node[mycircle, left =of v1v2] (v1) {$v_1$};
      \node[mycircle, below=of v3v4] (v4v3) {{ $v_4$},$v_3$};
      \node[mycircle, right=of v4v3] (v3v2) {{ $v_3$},$v_2$};
      \node[mycircle, fill = red, font = \huge, fill opacity = 0.4, right=of v3v4] (ax){$\textcolor{white}{\blacksquare}$};


    \foreach \i/\j/\txt/\p in {
      a1/v1/0.0/above,
      v1/v1v2/0.0/above,
      v1v2/v2v3/0.0/above,
      v2v3/v3v4/0.0/above,
      a1/v3/0.0/above,
      v3/v3v4/0.0/above,
      v3v4/v4v3/0.0/above,
      v4v3/v3v2/0.0/above,
      v3v2/ax/0.0/above,
      v3v4/ax/0.0/above}
       \draw [draw = black, myarrow] (\i) -- node[sloped,font=\small,\p] { } (\j);

    
\end{tikzpicture}} \\ \hfill  \\
    \hspace{0.24\linewidth} Order 1 \hfill Order 2 \hspace{0.26\linewidth}
    \caption{Example of a 1st order network (basic directed network) and the corresponding 2nd order network.
    Note that in the 1st order network, there is a potentially endless loop from $v_2 \rightarrowsmall v_3 \rightarrowsmall v_2$, which is not present in the 2nd order network.}
    \label{fig:hon-example}
\end{figure}

The state-space of all possible combinations of $k$ sequential nodes $\vec{v}\ord k$ can be constructed using ${\curly V}^k$, the $k^{th}$ power of the node set $\curly{V}$. 
Note that not all possible combinations will be present in the data: since observable combinations will be restricted by the $1^{st}$ order network topology, we define the observed set of higher order nodes 
$\curly V \ord k \subseteq {\curly V}^k$ where $\forall \vec{v}\ord k \in \curly V \ord k \exists \tilde S \in \tilde{\curly D}$ with $\vec{v} \ord k \in \tilde S$. 
In other words, $\curly V \ord k$ only contain higher order nodes of order $k$ that is observed in the data. Alternatively, one can think of the $k^{th}$ order nodes as the edges of the $(k-1)^{th}$ order network. We can then connect these higher order nodes using the same rules as one would use to construct a standard directed weighted network. In the example shown in figure \ref{fig:hon-example}, each order 1 edge between nodes that are not $*$ or $\dagger$ becomes a higher order node in the order 2 network. 

\subsection{Likelihood}
\label{subsec:likelihood}
In the previous subsection, we defined a method to generate a (multi-layer) higher order representation $\tilde S$ of any sequence $S$, as shown in equation (\ref{higher-order-sequence}). This representation naturally includes a correlation length defined by the order, as well as initial and final states, $*$ and $\dagger$, which correspond to how sequences start and end, respectively. The higher-order Markov transition probabilities between regular nodes are then reduced to the standard Markovian transition probabilities between higher order nodes, as shown in equation (\ref{higher-order-nodes}). This also allows for a convenient network representation akin to De Bruijn graphs, which we call a higher-order network, where the higher order correlations are sufficiently captured by dyadic edges between higher order nodes. This allows us to easily define the weighted directed adjacency matrix $A$, where each element $A_{u,v}$ counts the number of observed transitions $(u \rightarrowsmall v) \in \tilde S $ for each $ \tilde S \in \tilde{\curly D}$.  Crucially, however, we note that unlike traditional networks, $u$ and $v$ may also represent higher order nodes, as well as the initial and final states $*$ and $\dagger$ respectively. 


Having defined a methodology to generate a higher-order network and its adjacency matrix, we now turn to address how to create a stochastic generative model from the higher-order network. We can do this by defining a transition matrix $T$, where each entry $T_{u,v}$ represents the transition probability $P(v|u)$ from $u\rightarrowsmall v$.  The parameters $T_{u,v}$ can then be estimated by maximising the likelihood over the observed dataset $\curly D$. The formula for the likelihood of observing a single sequence $S = (v_1 \rightarrowsmall \cdots \rightarrowsmall v_{|S|})$ is given as
\begin{equation}
\begin{aligned}
    \curly L\Big(S=\big(v_1 \rightarrowsmall \cdots \rightarrowsmall v_{|S|}\big)\Big) = &
    {P (v_1|*) \times 
    \prod_{i=2}^k P (v_i| v_1 \rightarrowsmall \cdots \rightarrowsmall v_{i-1})} \\
    &\times \prod_{j=k+1}^{|S|} P(v_j|v_{j-k} \rightarrowsmall \cdots \rightarrowsmall v_{j-1}) \times P (\dagger|v_{|S|-k+1} \rightarrowsmall \cdots \rightarrowsmall v_{|S|}).
\end{aligned}
\end{equation}

\begin{example}[likelihood of a sequence] 
Taking the sequence $S = (v_1\rightarrowsmall v_2\rightarrowsmall v_3\rightarrowsmall v_4\rightarrowsmall v_5)$, and applying the likelihood defined above, the probability of observing the sequence under a $k=3$ higher-order network model is 

\begin{equation*}
\begin{aligned}
    \curly L(\tilde S) &=
    {P (v_1|*)} \times P(v_2|v_1) \times  P(v_3| v_1, v_2) \times P(v_4 |v_1, v_2, v_3) \times P(v_5 | v_1, v_2, v_3) \times P(\dagger |v_3, v_4, v_5) \\
    & = T_{*, v_1} \times T_{v_1, v_2} \times T_{(v_1,v_2), (v_2,v_3)} \times T_{(v_1, v_2, v_3), (v_2, v_3, v_4)} \times T_{(v_2, v_3, v_4), (v_3, v_4, v_5)} \times T_{(v_3, v_4, v_5), \dagger}.
\end{aligned}
\end{equation*}
\end{example}
The total likelihood across all sequences in the dataset is then simply the product of the likelihood of each sequence. We can then group factors of repeated observed transitions into powers to simplify the calculation. Thus, we can express the likelihood of observing the entire dataset $\curly{\tilde D}$ as \cite{gote2020predicting}

\begin{equation}
    \curly L \big(\curly{\tilde D} = \{\tilde S, \ldots \} \big) = \Bigg (\prod_{v \in \curly V} P(u|*)^{A_{*, u}} \Bigg) \times \Bigg( \prod_{i,j = 1}^{k} \prod_{v \in \curly V \ord i} \bigg [P(\dagger|v )^{A_{v, \dagger}} \times \prod_{w \in \curly V \ord{j}} P(w | v)^{A_{v,w}} \bigg] \Bigg),
    \label{eq:likelihood}
\end{equation}
where we recall that $A_{u,v}$ is the element of the weighted adjacency matrix, i.e. the number of observations of $(u \rightarrowsmall v)$.
We can then perform a maximum likelihood estimation of the parameters, namely the transition probabilities $P(v|u)$, using the method of Lagrange multipliers \cite{trench2012lagrange} under the constraints that $1 - \sum_{v}P(v|u) =0$ for all $u$. This yields the following result:

\begin{equation}
    P(v|u) = T_{u,v} = \frac{A_{u,v}}{\sum_{w} A_{u,w}},
\end{equation}
for any (higher order) nodes $u,v$, where the sum in the denominator is over all nodes $w$. 






\begin{figure}[h]
\centering
Sample dataset $\curly D = \big\{ (a, b, c, \ldots), \ldots \big\}$

\includegraphics[width = 0.9 \linewidth]{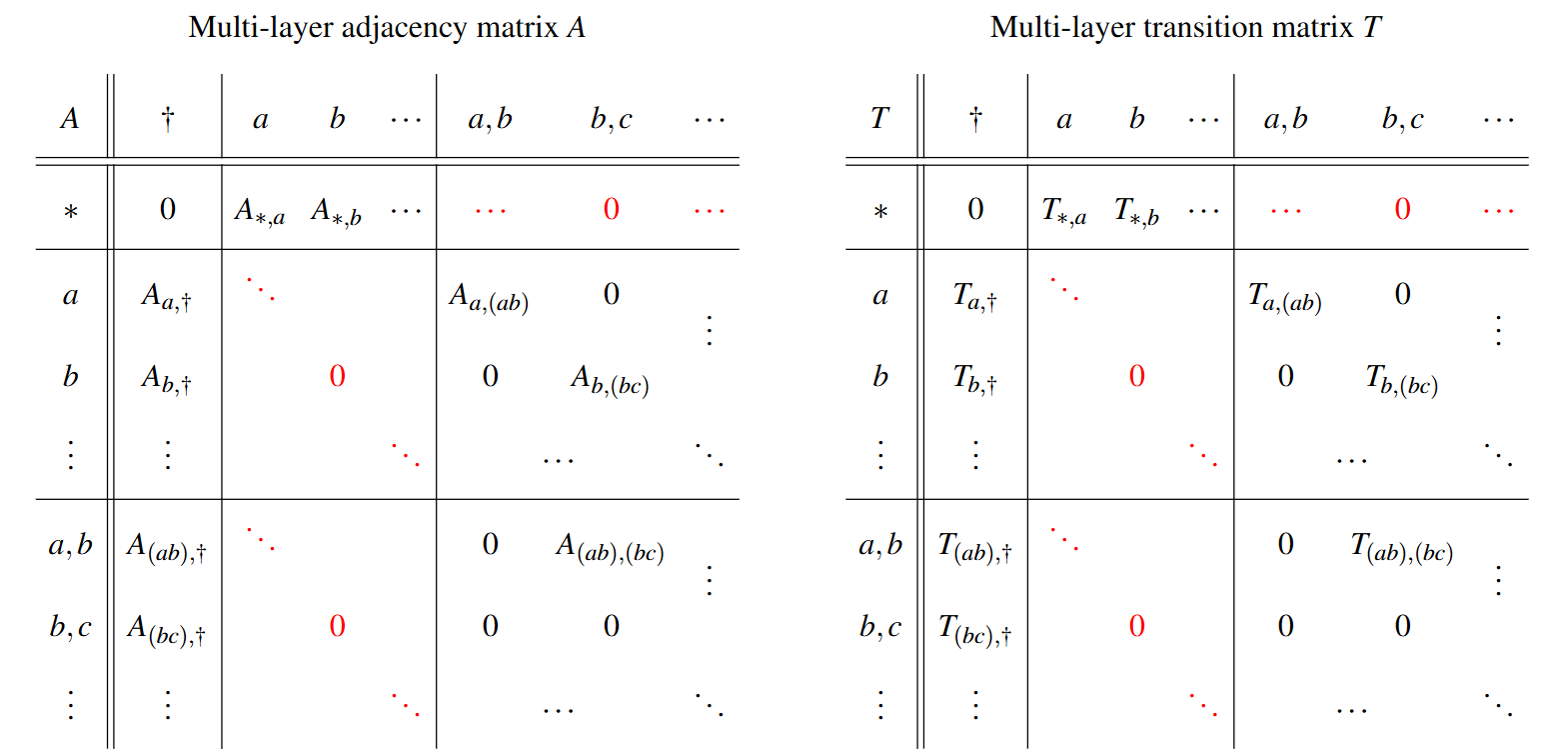}
    \caption{Example of an order $2$ adjacency matrix (left) and transition matrix (right) of an example dataset. The only transitions allowed are (1) from the starting state $*$ to order $1$ nodes, (2) from order $1$ to order $2$ nodes, (3) from order $2$ nodes to other order $2$ nodes, and (4) from any node to the state $\dagger$. All other transitions are prohibited, highlighted in red. For the transition matrix, each row is normalised such that all elements in the row sum to $1$.}
\label{fig:adj-trans-matrix}
\end{figure}



The higher order adjacency and transition matrix takes on properties of a multi-layer network adjacency matrix, where each layer only contains nodes of a specific order, up to a maximum order. An example is shown in figure~\ref{fig:adj-trans-matrix}. For a given maximum order $k$, the matrix elements are:
\begin{itemize}
    \item $\geq 0$ from the initial state $*$ to the $1^{st}$ order layer,
    \item $\geq 0$ from any $i^{th}$ order layer to the final state $\dagger$,
    \item $\geq 0$ from the $i^{th}$ order layer to $(i+1)^{th}$ order layer, up to $i = (k-1)$,
    \item $\geq 0$ entries from $k^{th}$ order nodes to other $k^{th}$ order nodes,
    \item $0$ in all other entries.
\end{itemize}

A potentially useful alternative and equivalent expression is to combine transitions from lower orders~$\leq k$ up to order $k$ (where possible), written as 

\begin{equation}
    {T_{*, v_1} \times 
    \prod_{i=2}^{k} T _{(v_1, \cdots, v_{i-1}), (v_{1}, \cdots, v_{i})}} = T'_{*, (v_{1}, \ldots, v_{k})},
\end{equation}
where $\tilde T $ is a structurally different transition matrix. While it is mathematically equivalent in terms of the likelihood function, it is computationally different as it reduces the number of redundant orders. Therefore, lower order transitions from initial state $*$ immediately jump to the highest order node, skipping potentially multiple intermediary transitions between the lowest and highest order nodes. 

\begin{example}[comparing a multi-layer transition to a highest-order transition] Consider the higher order sequence with $k=3$:
\begin{equation*}
    \tilde S = (
    * \underbrace{\xrightarrow{T_{*, v_1}}
    v_1 \xrightarrow{T_{v_1, (v_1, v_2)}}
    (v_1, v_2) \xrightarrow{T_{(v_1, v_2), (v_1,v_2,v_3)}}
    }_{T'_{*, (v_1,v_2,v_3)}}   
    (v_1,v_2,v_3) 
    \xrightarrow{T_{\cdots}} (v_2,v_3,v_4)
    \xrightarrow{T_{\cdots}} \cdots
    )
\end{equation*}
This can alternatively be represented as 
\begin{equation}
    S' = (* \xrightarrow{T'_{*, (v_1,v_2,v_3)}}
    (v_1,v_2,v_3) \xrightarrow{T'_{\cdots}}
    (v_2,v_3,v_4) \xrightarrow{T'_{\cdots}}
    \cdots).
\end{equation}
In this example, the number of steps (transitions) needed to go from $*$ to $(v_1, v_2, v_3)$ in $\tilde S$ is 3 steps, while in $S'$ only 1 step is required. When generating new sequences using a stochastic process such as a random walk, each step computationally requires the generation of a random number. Reducing the number of these random choice operations can improve the efficiency of simulating new sequences. The highest-order transition matrix $T'$ does this while maintaining the exact same probability distribution of sequences as the multi-layer transition matrix $T$. 

\end{example}
\arraycolsep=4pt\def\arraystretch{2}
\begin{figure}[h!]
\centering
\text{Sample dataset $\curly D = \big\{ (a, b, c, \ldots), \ldots \big\}$}\\
    
    
    
    
    
    
\includegraphics[width = 0.5\linewidth]{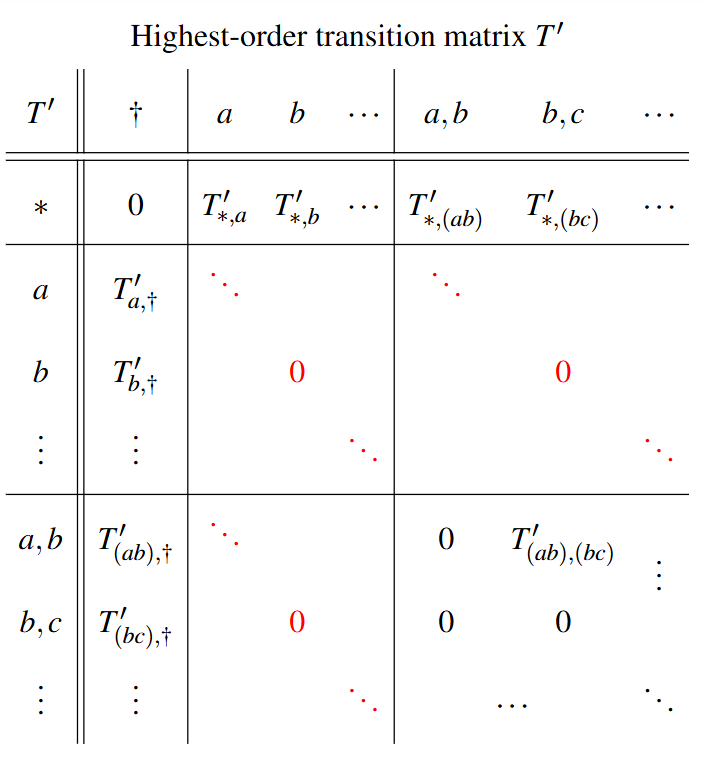}
\caption{Example of a highest order 2 transition matrix of the sample dataset, compared to the multi-layer transition matrix in figure \ref{fig:adj-trans-matrix}. In contrast to the multi-layer matrix approach, the highest order matrix immediately transitions from the initial state $*$ to the highest order nodes (where possible). Transitions from $*$ to a node with order $<k$ is forbidden, unless the node itself is an entire sequence.}
\label{fig:transition-matrix-2}
\end{figure}

\subsection{Expected Sequence Lengths}
As each individual sequence must terminate at the final state $\dagger$, we can think of $\dagger$ as an absorbing state in a Markov chain, with all other state being transient. However, the multi-layer transition matrix $T$, as defined in the previous subsection, is not a transition matrix of an absorbing Markov chain~\cite{grinstead2012introduction}. Nonetheless, it bears some similarities that we can utilise. For a higher-order network containing $n$ nodes in addition to the $*$ and $\dagger$ states, the transition matrix $T$ of this network can be partitioned in a similar manner to the canonical form
\begin{equation}
T = 
    \left(\begin{array}{c|cc}
        T_{*,\dagger} & T_{*,a} & T_{*,\cdots} \\ \hline
        T_{a, \dagger} & T_{a,a} & \cdots \\
        T_{\cdots, \dagger} & \vdots & \ddots \\        
    \end{array}\right) =
    \left(\begin{array}{c|c}
        T_{*,\dagger} & T_{*, Q} \\ \hline
        T_{Q, \dagger} & Q
    \end{array}\right),
\end{equation}
where $Q$ is the $n \times n$ sub-matrix of $T$ that contains all the transition probabilities between transient states, which in this case are all the (higher order) node-to-node transitions. $T_{*, Q}$ is the $1\times n$ matrix containing all the initial transition probabilities,  $T_{Q, \dagger}$ is the $n \times 1$ matrix containing all the final transition probabilities, and $T_{*,\dagger} = 0$. If we replace $T_{*, Q} = 0_{1 \times n}$ and $T_{*,\dagger}=1$, essentially turning $\dagger$ into a true absorbing state where transitions out of it cannot occur, we obtain the canonical form of the absorbing Markov chain
\begin{equation*}
    \left(\begin{array}{c|c}
        1 & 0_{1\times n} \\ \hline
        T_{Q, \dagger} & Q
    \end{array}\right).
\end{equation*}
The expected sequence length $\mathbb E(L)$ is then given by \cite{grinstead2012introduction} 
\begin{equation}
    \mathbb E(L) = \sum_{i,j} [(T_{*, Q})^T (I_{n\times n} -  Q)^{-1} ]_{i,j},
\label{eq:expected-sequence-length}
\end{equation}
where $I_{n\times n}$ is the $n\times n$ identity matrix. Notably, this equation only holds for the multi-layer transition matrix $T$.

\subsection{Overview of Model Selection Techniques}
 
In the previous subsections, we describe our methodology to generate a higher-order network model at any given order from a dataset of sequences. This yields a set of possible models at various orders, which naturally raises a common model selection problem: which order provides the best model? Broadly speaking, this typically corresponds to selecting the model that has maximum explanatory power and minimum  complexity. Model selection techniques provide statistically rigorous measures to discriminate between models, and these techniques have been widely implemented in many fields. However, the higher-order network model selection problem introduces some undesirable features, especially regarding the model complexity. The number of parameters required to specify the model increases exponentially with the order. This means that the measures defined by various information criteria may not be close to their asymptotic limits, and may therefore be prone to inaccuracies \cite{ding2018modelselection}. As there is no single `best' information measure for this task, we instead take a consensus approach by applying multiple information measures. Prior work on \verb|PathPy| \cite{scholtes2017network} and MOGen \cite{gote2020predicting} used the likelihood ratio test \cite{wilks1938large} and the Akaike Information Criterion (AIC) \cite{aic}. In addition to applying these two methods, we also compare them with the Bayesian Information Criterion (BIC) \cite{bic, schwarz1978estimating}. 

There are two key factors that play an important role in all three measures: the likelihood function in equation~\ref{eq:likelihood} evaluated at its maximum $\curly{{\hat L}}$, and the number of degrees of freedom $\delta$. Intuitively, the 
likelihood function measures the goodness-of-fit of the model, while the number of degrees of freedom acts as a penalty term for increasing model complexity. With higher-order networks, the maximum likelihood, as a function of the order, monotonically \textit{increases} in the range $(0,1$]. Its logarithm is negative, so~$-\ln (\curly{\hat{L}})$ is therefore a monotonically \textit{decreasing} function in the range $(0, \infty)$. On the other hand, the number of degrees of freedom $\delta$ monotonically \textit{increases} with order. 

\begin{figure}
    \centering
    \includegraphics[width=0.5\linewidth]{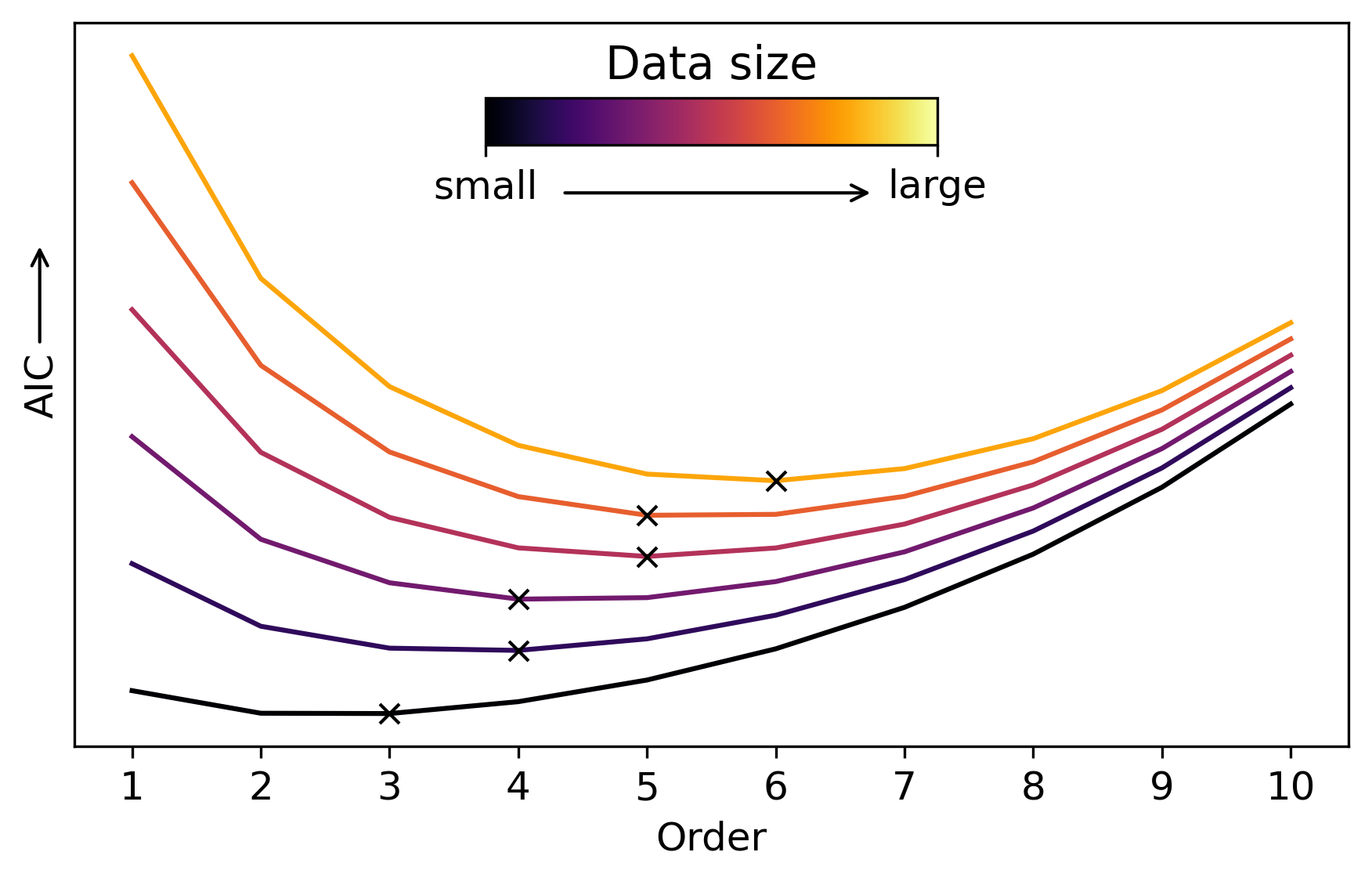}
    \caption{Illustration showing the effects of data size and order on AIC measure, for some arbitrary data size. The order that minimises the AIC measure for each data size is marked by a black $\times$. The trend shows that as the data size increases, the order that minimises the AIC measure increases as well.}
    \label{fig:aic-datasize}
\end{figure}

In the AIC formulation, the optimal model is the model that minimises the following measure

\begin{equation}
    \text{AIC} = 2\delta - 2 \ln(\hat{\mathcal L}),
\end{equation}
where the penalty term is $2\delta$. This safeguards against higher orders where the number of parameters grows approximately exponentially. Note that while the $-\ln (\curly{\hat{L}})$ term scales with the size of the data (as it will increase the number of terms in the product of the likelihood function), the penalty term generally does not. Additionally, the $-\ln (\curly{\hat{L}})$ term scales \textit{faster} at lower orders. Put together, as more data is added to the observation, the $-\ln(\curly{\hat L})$ at lower orders increases faster than at higher orders, while the penalty term $2\delta$ generally remains constant. Thus, increasing the number of observations in the dataset can allow higher order models to be selected. In the limit where the data size $|\curly D| \rightarrow \infty$, AIC will tend to select the highest order possible, i.e. $k = \text{max}(\{|S|: S \in \curly D\})$; in practice, this is less of a concern as the increase in data size required to observe this effect is typically several orders of magnitude. An illustration of this effect is shown in figure \ref{fig:aic-datasize}.

The BIC measure is similar to AIC in that it uses the same goodness-of-fit term $-2 \ln(\hat{\mathcal{L}})$, but has a stricter penalty term that depends on the size of the dataset $|\curly D|$:

\begin{equation}
    \text{BIC} = \delta\ln(|\curly D|) - 2 \ln(\hat{\mathcal{L}}),
\end{equation}
where the penalty term is $\delta \ln(|\curly D|)$. Notably, in contrast to the AIC measure, the penalty term scales with the data size as well, and is therefore less sensitive to the number of observations. As $\ln(|\curly D|) >2$ in most cases, the penalty term is much larger in BIC than AIC. Further discussion of the use and comparison between AIC and BIC can be found in \cite{ding2018modelselection}. 

The last criterion used is the likelihood ratio test between two higher-order models $\curly M_{k}$ and $\curly M_{k+1}$ of orders $k$ and $k+1$ respectively \cite{scholtes2017network}. Given the likelihood ratio $\Lambda$,  Wilks' theorem states that  $-2\ln (\Lambda)$ asymptotically converges to a $\chi^2$ distribution under the null hypothesis $H_{0}$ that the two models $\curly M_{k}$ and $\curly M_{k+1}$ are not statistically distinguishable \cite{wilks1938large}:

\begin{equation}
    \lim_{d \to \infty} -2\ln\bigg(\frac{\curly{\hat{L}}(\curly M_{k})}{\curly{\hat{L}}(\curly M_{k+1})} \bigg) \sim \chi^2_{\delta_{k+1} - \delta_{k}}
\end{equation}
We can then obtain a $p$-value for accepting the alternate hypothesis $H_{alt}$, which corresponds to selecting the higher order model $\curly M_{k+1}$ over $\curly M_{k}$. In this paper, we choose $p = 0.05$, so that higher order models are only chosen if they have a significantly higher maximised likelihood with respect to the increased number of parameters. The likelihood ratio test is then performed iteratively,  comparing consecutive orders until the null hypothesis is accepted. 

\subsection{Macroscale and mesoscale metrics to evaluate model performance}

The model selection techniques discussed in the preceding subsection  may not adequately fully evaluate how well a model represents the system that generated the data. Notably, the measures used in these techniques are relative measures between models. A challenge therefore is to identify whether the models are objectively good. The selected best model may still be a poor representation of the system if all other models are worse. Therefore, unless a good model is in the pool of potential models, or some mechanistic knowledge of the system is known \textit{a priori}, it remains an issue that needs to be addressed. As such, we consider other measures from the perspective of identifying a model that can generate sequences that reflect certain features of the data that we used to parameterise the model. To do this, one has to define what `agreement' between data and model output looks like. There are a number of options available for this.


First, we propose investigating macroscale statistics such as sequence lengths and their distributions, and require that the sequence length distributions from the data and those generated by the models are similar. Good agreement at this scale guarantees that sequences of varying lengths are captured correctly. We evaluate the model's ability to reproduce the observed sequence length distribution via a random walking process on the transition matrix using the Kolmogorov-Smirnov (KS) test. Using models of different order, we simulate multiple stochastic walks and compare statistics from these simulations to those derived directly from the data. The KS test is a non-parametric statistical test to evaluate whether two samples could be randomly drawn from the same underlying distribution. However, agreement with respect to the KS test does not imply that sequences visit nodes in the right order. Hence, additional measures of agreement between data and models are required. We also propose considering the frequency of different motifs over $3$ and $4$ nodes, measured from both the data and from simulated sequences generated by the model. This is then evaluated by comparing the residuals between the data and model simulations. 

Based on previous work, one might expect that the best agreement between data and model will be achieved at the optimal order \cite{chawla2016representing, scholtes2017network, gote2020predicting}. We extend this to also study the change in accuracy of these macro- and meso-scale metrics as a function of the order.  

\section{Results} \label{Results}

We tested our model by analysing several datasets:
\begin{itemize}
    \item Bike share services \cite{fournier2016spmf, labikesharemetro, tfl2023bikejourney}: data contains the movement of bikes between docking stations recorded over a period of time. Two datasets were obtained from different cities, one in Los Angeles and the other in London.
    \begin{itemize}
        \item L.A. \cite{labikesharemetro}: recorded 21,078 rides from 2016-07-07 to 2017-03-31.
        \item London \cite{tfl2023bikejourney}: recorded 44,543 rides from 2023-01-16 to 2023-01-22.
    \end{itemize}
    \item User clickstream data from MSNBC \cite{fournier2016spmf}: recorded 989,818 sequences of pages visited by users on the MSNBC website.
    \item US flight itinerary survey \cite{bts2023odsurvey}: recorded 8,570,568 flight itineraries indicating the states visited in the US.
    \item Taxis in Porto \cite{taxiECMLPKDD2015}: data contains the pickup and drop-off locations and time-stamps of taxi rides in Porto, Portugal. A total of 302,688 sequences were recorded over the period from 2013-07-01 to 2014-07-01. 
    \item Proprietary global shipping data: records the movement of ships between $26$ geographical regions around the globe from 2016-01 to 2020-01, for a total of $9,505$ sequences. Ships are split into 5 types, which correspond to the class of the ship based on their maximum tonnage. 
\end{itemize}
All datasets apart from the shipping dataset, which is subject to a Non Disclosure Agreement, are accessible from open online sources.  

We created higher-order network models of the data using the method described in the previous section, exploring various orders from $1$ up to $4$. Orders beyond order $4$ were avoided not only because they were computationally taxing, but they also tended to display over-fitting tendencies. This is due to the exponentially growing number of parameters fitted to the model, where the upper bound is given by $\sum_{i=1}^{k} |\curly V|^i \approx |\curly V|^k$.

These models were then evaluated against the data by simulating new random sequences via a random walk process, described by the following algorithm:

\begin{enumerate}
    \item Initialise an empty sequence list \verb|S|.
    \item Set the current node \verb|u| as the initial state *.
    \item Choose the next node \verb|v| at random with probabilities equal to $T_{u,v}$ (or $T'_{u,v})$.
    \item Append node \verb|u| to \verb|S|.
    \item Set node \verb|u| = \verb|v|.
    \item Repeat steps (3-5) until \verb|v=†| is chosen.
    \item Terminate and record \verb|S|.
\end{enumerate}
For all datasets, unless stated otherwise, we simulated $1,000,000$ random walk sequences per order. Measurements on the simulated sequences provide the expected model outcome for the measurement.

\subsection{Model Selection of Higher-Order Networks}
Introducing higher order correlations into networks naturally gives rise to the question: what is the `right' correlation length or order? In the previous section, we discussed a variety of model selection techniques, aiming to identify the optimal order for the model. However, higher-order networks introduce an added layer of complication due to the vastly different sizes of parameter spaces at different orders. This poses a meta issue where the `best' model selection technique itself is not definitively known. We propose taking a consensus approach by applying and comparing 3 common techniques: the Akaike Information Criterion~\cite{aic}, the Bayesian Information Criterion~\cite{bic}, and the likelihood ratio test~\cite{wilks1938large}. The results are shown in table~\ref{tab:model-selections}.



\begin{table}[h!]
\centering
\def\arraystretch{1.2}
\begin{tabular}{|c|c|c|c|}
    \hline 
    Data & AIC& BIC& Likelihood Ratio\\
    \hline
    Ship Type 1& 2& 1& 2\\
    Ship Type 2& 1& 1& 2\\
    Ship Type 3& 2& 1& 2\\
    Ship Type 4& 2& 1& 2\\
    Ship Type 5& 2& 1& 2\\
    MSNBC & 3& 2& 3\\
    Bike (L.A.) & 1& 1& 1\\
    Bike (Lon) & 1& 1&1\\
    US Flights & 3& 2& 3\\
    Taxi& 3& 2&3\\
    \hline
    \end{tabular}
    \caption{Optimal order selected by 3 different criteria: the Akaike Information Criterion (AIC), Bayesian Information Criterion (BIC), and the likelihood ratio test via Wilks' theorem. The general trend shows that AIC and the likelihood ratio agrees most of the time, while BIC tends to select 1 order lower than either AIC or the likelihood ratio test.}
    \label{tab:model-selections}
\end{table}

Table~\ref{tab:model-selections} displays the optimal order obtained via the 3 model selection techniques. The trend shows that AIC and the likelihood ratio test agrees for most datasets, with the exception of Ship Type 2. Meanwhile, BIC tends to select 1 order lower than AIC, which is an expected result due to the harsher penalty term in BIC. However, these information measures do not provide an absolute measure of how well the model captures the data; they mainly provide a relative comparison between models. Thus, we investigate other metrics in the following subsections at various orders in addition to the optimal order. 

\subsection{Sequence Length Distribution}

\begin{figure}
    \centering
    \includegraphics[width = 0.75 \linewidth]{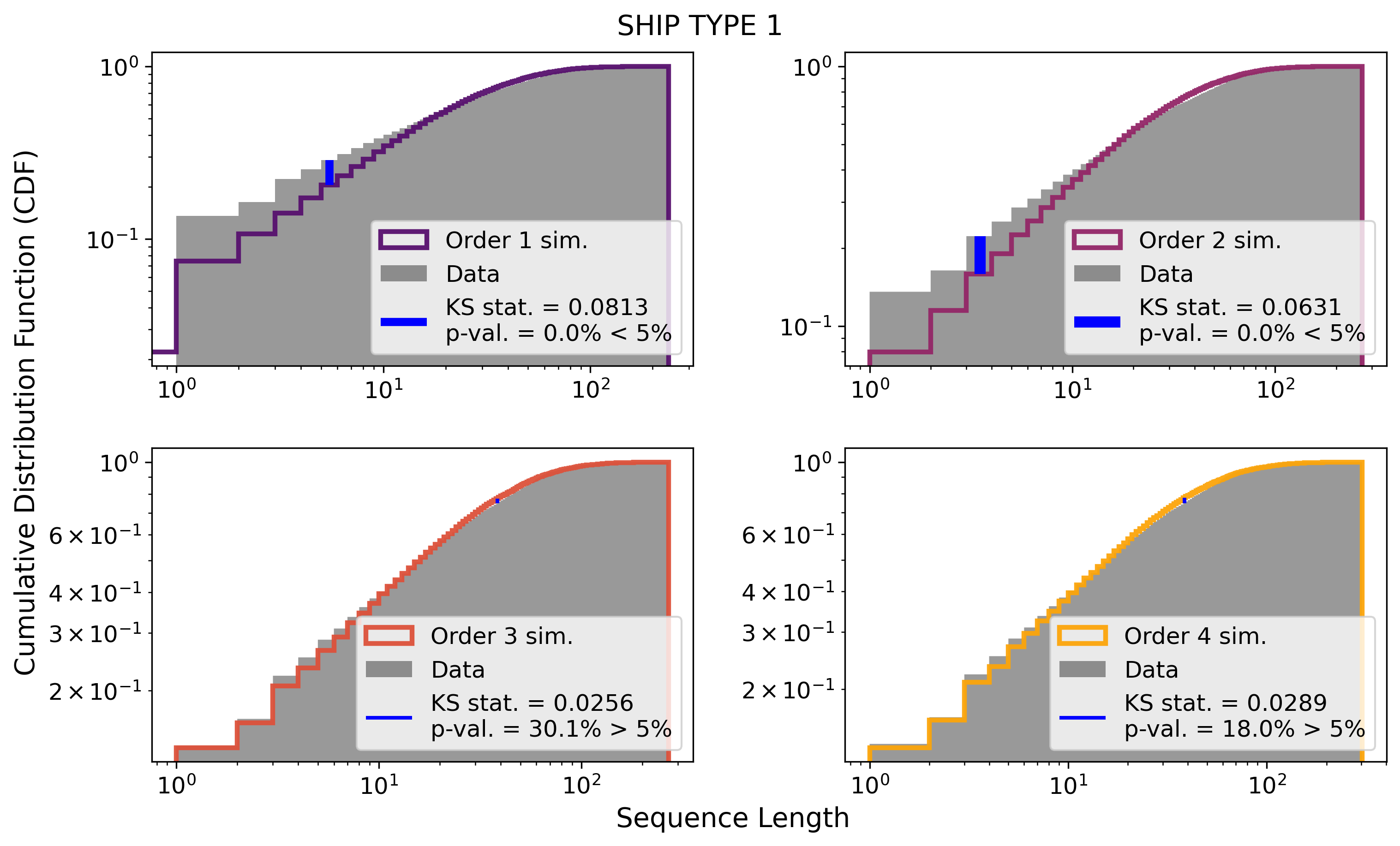}

    \caption{Example plot of the empirical cumulative distribution function (CDF) from observed and simulated sequence lengths, from Ship Type 1. The magnitude and location of the KS statistic and $p$-value are shown as well. The results show a general improvement in the KS-statistic as the order increases. The $p$-value is within an acceptable range at order 3. The expected sequence lengths obtained from equation~\ref{eq:expected-sequence-length} gives a value of $25.09...$ for all orders, which agrees with the data.}
    \label{fig:ks-example}
\end{figure}

Real world data often comes in the form of finite variable length sequences. Finitude is a natural constraint of real world systems, while variability arises from other random factors. The inclusion of the initial state and final state offers a framework for initiating a random walk process that self-terminates, in order to better emulate how real-world agents or processes enter and leave the network. Biases in where/how sequences start may significantly impact its evolution, particularly for short sequences. Similarly, biases in where/how sequences terminate may dramatically alter the final length of the distribution. 
A simple initial test used to validate that the model is working as intended is to calculate the expected sequence length based on the transition matrix, as shown in equation~\ref{eq:expected-sequence-length}. Our results (see the caption of figure~\ref{fig:ks-example}) present an accurate match with the data. This provides confidence that the model is generating sequences that are, at the very least, comparable with the data. However, averaging sequence lengths discards a lot of information about the variability in the sequence lengths. We therefore focus on comparing the \textit{distribution} of sequence lengths in order to better understand the similarities and differences between the simulations and data.


In order to compare the sequence length distributions between the data and random walk simulations, we employ the Kolmogorov-Smirnov (KS) test \cite{lewis2017simulation}. The KS statistic $D_{KS}$ is measured as the maximum absolute distance between two cumulative distribution functions (CDFs). Naturally, $D_{KS} \leq 1$. A $p$-value can be then obtained, under the null hypothesis that both distributions can be explained as being generated by the same underlying distribution. This is a particularly strong condition for accepting the null hypothesis. The KS test, however, does not have any penalty term for the complexity of the model, and therefore we need to be cautious with regards to over-fitting. Figure~\ref{fig:ks-example} shows an example of the two CDFs at order $1$ to $4$, as well as the location of the KS statistic and the $p$-value, from the Ship Type 1 dataset. We present our findings in table~\ref{tab:ks-test}, showing the order at which the KS-test shows acceptance (or none, if no acceptance was found between order 1 and 4), as well as the KS statistic and the $p$-value at that order. 


\begin{table}[h!]
\def\arraystretch{1.2}
\centering
\begin{tabular}{|c|ccc|}
    \hline
    \multicolumn{4}{|c|}{Walk Length Distribution}\\
    \hline
    Data & Selected Order&KS Statistic &$p$-value\\
    \hline
    Ship Type 1 & 3& 0.03&30.1\%\\ 
    Ship Type 2 & 3& 0.07&9.0\%\\  
    Ship Type 3 & None$^1$ & -&-\\
    Ship Type 4 & 3& 0.02&43.2\%\\
    Ship Type 5& None$^1$& -&-\\
    MSNBC & None$^{1}$ & -&-\\
    Bike (L.A.) & 2& 0.008&18.7\%\\
    Bike (Lon) & 3& 0.006&93.5\%\\
    US Flights & 4$^2$& 0.001&96.5\%\\
    TAXI&None$^1$ & -&-\\ 
    \hline
\end{tabular}
\caption{Selected order is the minimum order that passes the KS test (i.e. $p \geq 5\%)$, up to a maximum order of $4$, across all datasets. This differs from the optimal orders obtained from model selection techniques as shown in table \ref{tab:model-selections}.  $^1$A $p$-value $\geq 5\%$ was not observed at any order, which suggests that there is a more complex behaviour behind the ship movements. $^2$The high $p$-value suggests that the model is over-fitting, which seems to be the case for the Flight data as majority of itineraries contain $\leq 4$ flights, and therefore at order $4$ many sequences are entirely encoded in the network.}

\label{tab:ks-test}
\end{table}
Our findings suggest that for certain datasets (some Ship Types, Bikes and Flights) the order required to reproduce the path length distribution observed in the data is generally higher than that estimated by optimal order selections. The bike datasets show the lowest order for acceptance, which is both expected and surprising. Bikes themselves do not have any agency in their movement, and therefore we do not expect any temporal correlations. However, this would suggest that the KS test should be accepted at order $1$, rather than order $2$. One possible explanation for this would be that certain users are docking and re-using the same bikes in succession, possibly due to time limits on the riding time between docks. For the Flight dataset, the main discrepancy occurs for sequences of length $4$, which corresponds to double-layover return-flights. Due to this, at order $3$ the KS statistic observed at sequence length $4$ is too large to be acceptable, but at order $4$ the KS statistic is too small due to the model over-fitting. In the MSNBC data, our findings suggests that the increase in order is not warranted as the difference in accuracy of simulated versus observed data is not significant (at least up to the $4^{th}$ order). A possible explanation is that the mechanism underlying the sequence termination is not related to its past; for example, agents looking for specific pages might leave the website after finding the page they want. Similar ideas can be applied to other datasets where no order is accepted from the KS test. A potential factor to consider is that sequences may be truncated by external factors related to the data collection or processing itself. This effect is suspected to be at play in the London Bike and Taxi datasets.


\subsection{Motifs}
\subsubsection{Observation and Counts}
Sequence length distributions alone are not a conclusive indicator of a model's accuracy, since the right sequence length does not guarantee that paths are traversing nodes in the right order. Motifs provide another method of analyzing smaller scale features of the network. The expected fractions $\langle F \rangle$ of motifs under some higher-order network model can be estimated using random walk simulations on the network. By comparing the simulations to the data, we can evaluate how well the real-world motif dynamics are reflected in the model. First, we define $\mathbb M_l$ to be the set of $l$-hop motifs. As we can only compare motifs of the same length together, we will drop the subscript $l$. For each motif $M \in \mathbb M$, the observed count of $M$ in a sequence set $\curly D$ (which may be real or simulated) is $N_{M, \curly D}$. The expected fraction $\langle F \rangle_M$ of motif $M$ from the model can be estimated from

\begin{equation}
\begin{aligned}
    \langle F \rangle _{M} = \frac{N_{M, \curly D_{sim}}}{N_{\mathbb M, \curly D_{sim}}}, \;& \text{where } N_{\mathbb M, \curly D_{sim}} = \sum_{M \in \mathbb{M}} N_{M, \curly D_{sim}}, \\
    & \text{and } 0 \leq \langle F \rangle_M \leq 1
\end{aligned}
\end{equation}
and $\curly D_{sim}$ is a set of simulated sequences generated by the model. To compare the model to data $\curly D_{data}$, we multiply the expected fraction of each motif by the total number of motifs in the data:
\begin{equation}
    \langle N \rangle_M = \langle F \rangle_M \times N_{\mathbb M, \curly D_{data}}    
\end{equation}
This is to account for the discrepancies in the absolute counts in the data versus simulations.
Additionally, we can obtain an estimate of the variance $\sigma^2_M$ through the variance of the motif counts within each sequence itself. The studentized residual $R_M$ is then calculated as~\cite{fieberg2024statistics} 
\begin{equation}
    R^2_M = \frac{(N_{M,\curly D_{data}} - \langle N \rangle_M)^2}{\sigma^2_M}
    \label{eq:residuals}
\end{equation}

We evaluate motifs using several measures, including (1) absolute observed vs expected counts, (2) percentage difference between observed and expected counts, and (3) a studentized residual measure of goodness-of-fit. The absolute counts and percentage differences provide us with a more intuitive grasp of how close the expected motif counts from the model are compared to the observations. However, they do not provide us with an evaluation of whether the observations lie within an expected range of our model's predictions. The studentized residual gives us an approximate measure of the statistical `goodness-of-fit' of the model compared to the data. 
We begin first by observing the direct counts from the data compared to the simulations, where it is much easier to intuitively gauge which motifs are well represented and which are not. 

\begin{figure}
    \centering
    \includegraphics[width = 0.49 \textwidth]{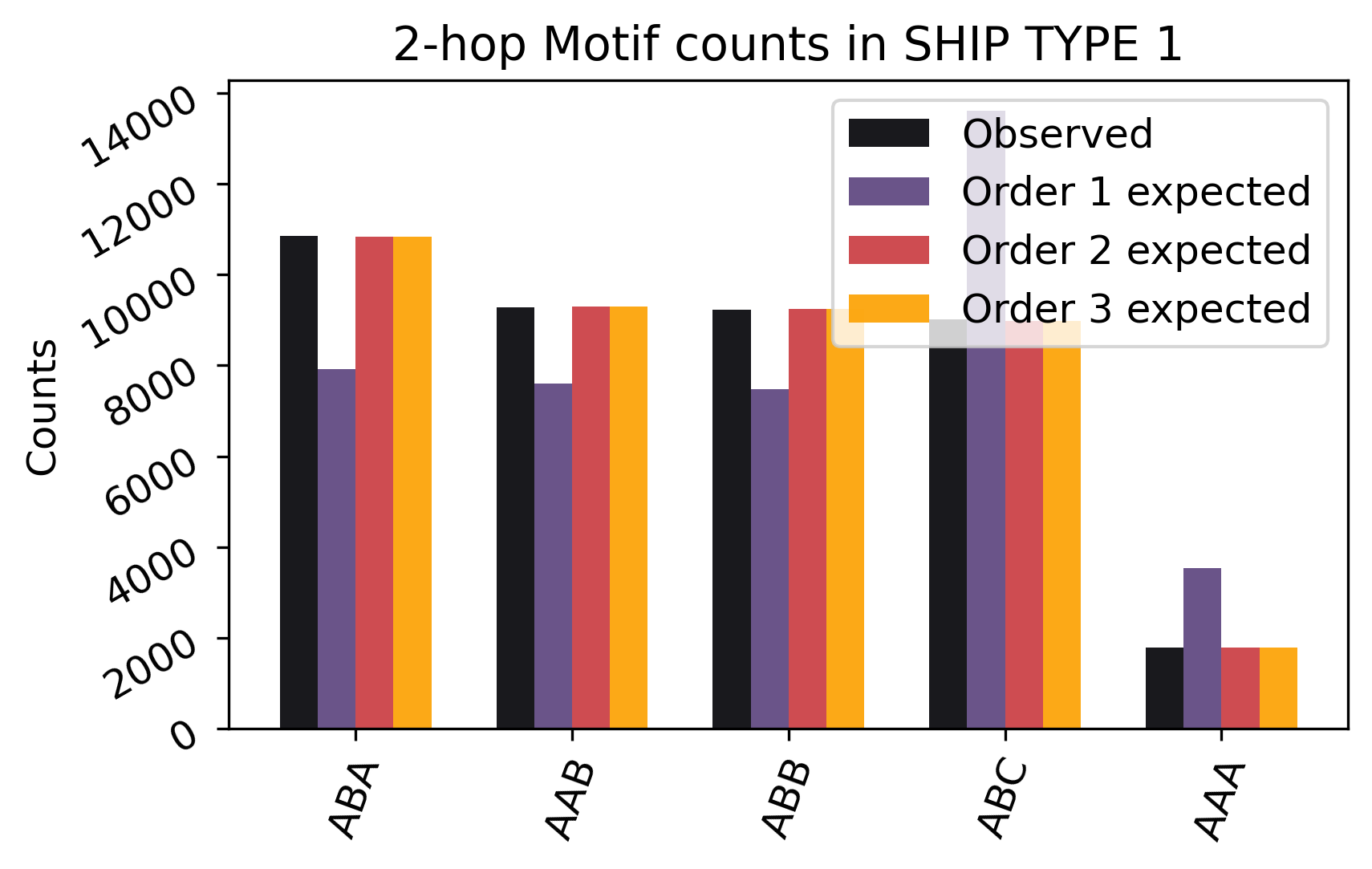}
    \includegraphics[width = 0.49 \textwidth]{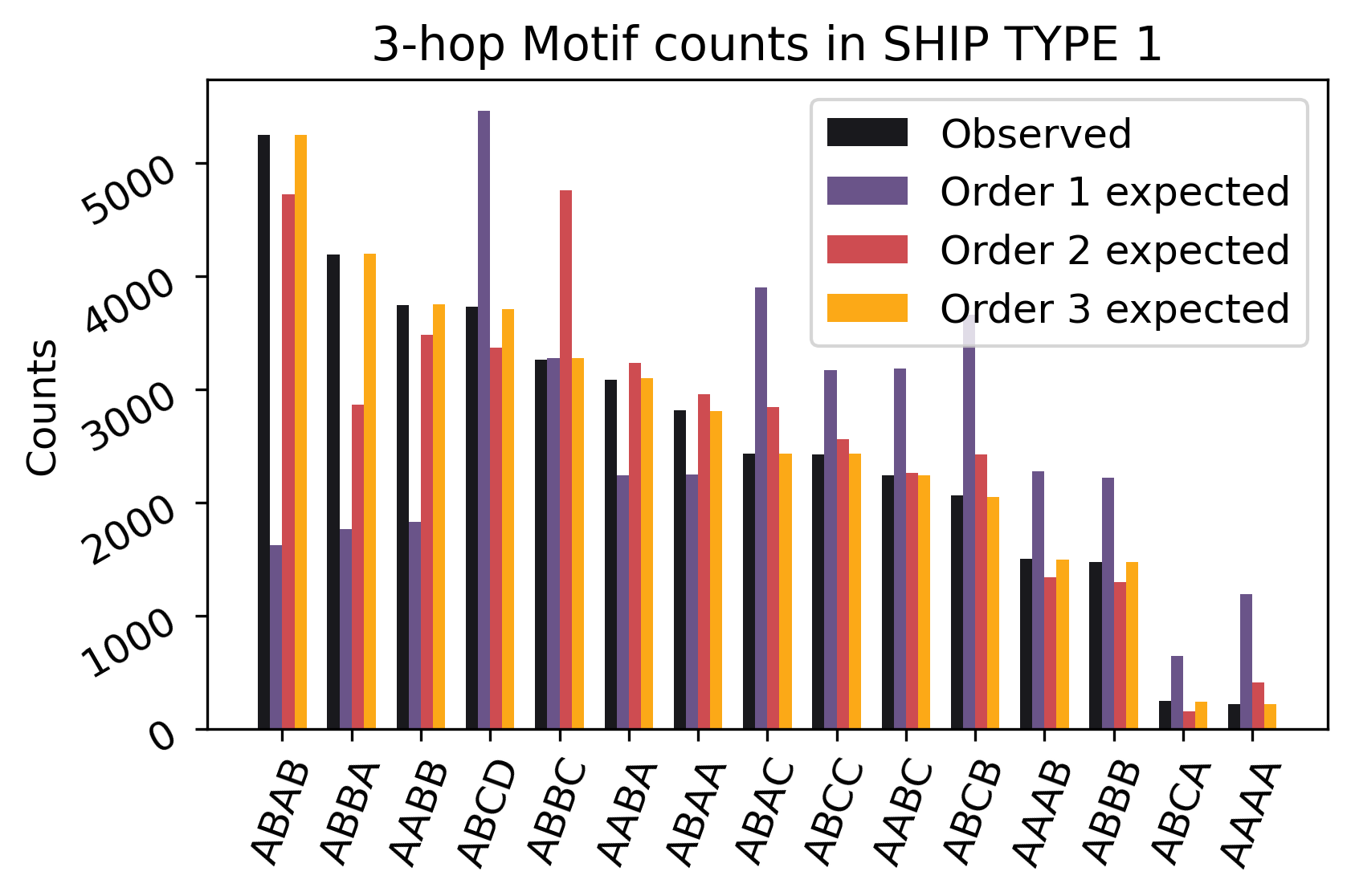}
    \caption{(Left): Counts of $2$-hop motifs from the data and from simulations on higher-order networks of order $1,2,3$. We observed agreement at order $2$ (pink) and above, which is expected. At order $1$ (purple), the $AAA$ motif and $ABC$ motif are underrepresented while all other motifs are overrepresented. (Right): Counts of $3$-hop motifs from the data and from simulations on higher-order networks of order $1,2,3$. Similarly, we observe agreement at order $3$ (orange). At order $2$ (pink) we observe relatively good representation of the observed motif counts for all $3$-hop motifs except $ABBA$ and $ABBC$.}
    \label{fig:motif-count-example}
\end{figure}

In Figure \ref{fig:motif-count-example}, we show the observed count of $2$- and $3$-hop motifs in one Ship Type dataset. As expected, the order 2 network is sufficient to reproduce the $2$-hop motif (e.g $ABA$) distribution in simulations. More generally, for $H$-hop motifs, an order $H$ network is sufficient to reproduce the observed motif distribution. This is because $H$-hop motifs can directly be mapped from the edges in the order $H$ network. A more interesting point of investigation is to look at the order $(H-1)$ HO network at the motif distribution compared to the observed data. In this paper, to avoid the large sample space of longer-hop motifs, we focus primarily on comparing how a $2nd$ order network performs to a $3rd$ order network. However, the analysis here can be generalised to longer motifs, provided that there is a reason for extending to higher orders.

For $3$-hop motifs, we can see that at order $2$, many motifs are well represented by the model compared to the data, with the main exceptions being the motifs $ABBA$ and $ABBC$. The $ABBA$ motif is underrepresented in the data, while $ABBC$ is overrepresented. Intuitively, both of these motifs can be seen as an $ABB$ motif chained with an $AAB$ motif. The order $2$ model correctly captures the frequency of both of $2$-hop motifs; however, it lacks the correlation length to `choose' the right $AAB$ motif such that it correctly predicts the frequency of $ABBA$ and $ABBC$ motif. This suggests that $ABBA$ and $ABBC$ are `genuine' $3$-hop motifs, in that they cannot be directly explained as a random chain of two $2$-hop motif with based on $2^{nd}$ order probabilities. In this dataset and in general, the genuine $3$-hop correlations correlations are dominant in only a few specific motifs. In certain cases, we may be able to attribute such motif patterns to an expected real-world behaviour as well. We can further examine which $3$-hop motifs are the most over/under-represented compared to the $2nd$ order network by observing the difference in observed and expected counts, and ranking them accordingly. 

\begin{figure}
    \centering
    \includegraphics[width = 0.49\linewidth]{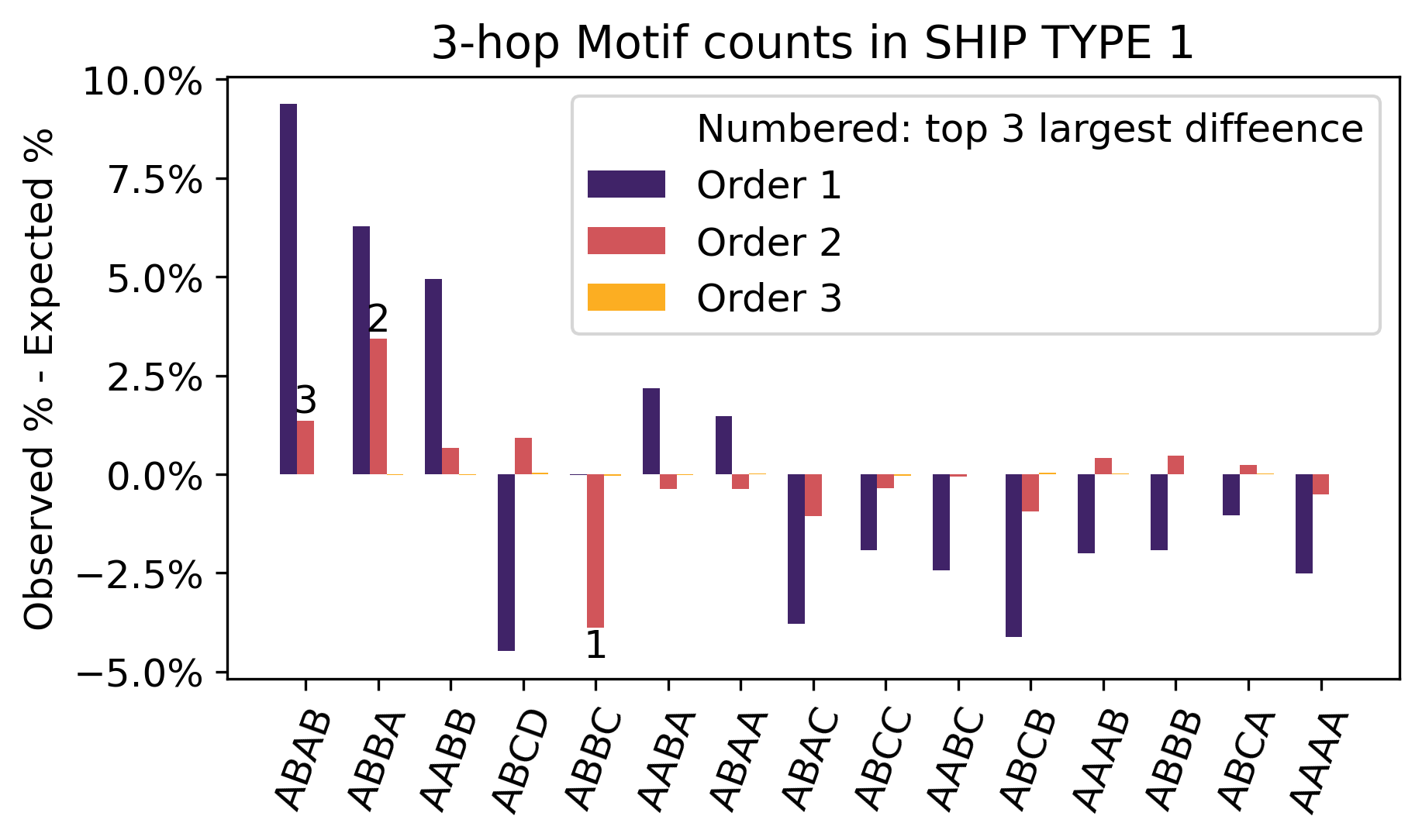}
    \includegraphics[width = 0.49\linewidth]{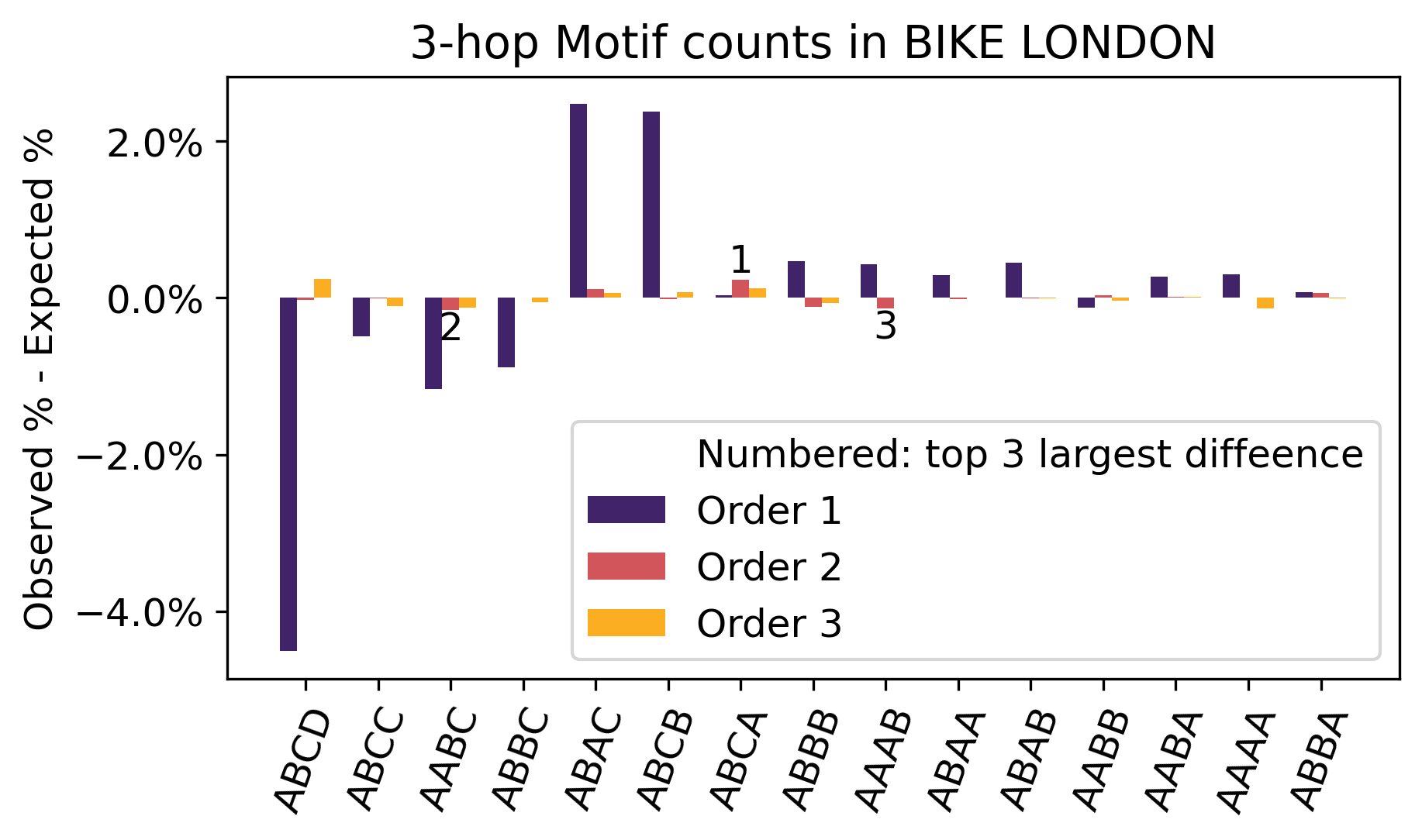}
    
    \caption{Difference between the observed and expected percentages in the number of $3$-hop motifs at various orders. The 3 largest percentage differences at order $2$ (pink) are indicated by the numbers (rank) above the bars. (Left): Data shown for Ship Type 1, the motifs $ABBC$, $ABBA$ show the largest difference, constituting over 2.5\% of the total motif counts each, followed by $ABAB$. (Right): Data shown for Bike London, the motifs $ABCA$, $AABC$ and $AAAB$ are the worst performing motifs; however, the deviation is extremely small, and therefore is not significant.}
    \label{fig:total-percentage}
\end{figure}

The results from figure~\ref{fig:total-percentage} show that for Ship Type 1, the motifs $ABBC$ and $ABBA$ are the worst represented motifs. Additionally, majority of the other 12 motifs are under a $2\%$ difference, which suggests that a $2^{nd}$  order network captures most of the sequential correlations in the data very well. This trend is generally present in other datasets as well: most motifs show a small deviation between simulations and data, except for a few large deviations. Finally, as expected, we can see that for most motifs, the $2^{nd}$ order network performs better than the $1^{st}$ order network.  

\begin{table}[h!]
\def\arraystretch{1.2}
\centering
\begin{tabular}{|c|c|c|c|}
    \hline 
     & \multicolumn{3}{c|}{Worst Represented 3-Hop Motifs ($\Delta$ \%)} \\
     \hline 
    Data & 1& 2& 3\\
    \hline
    Ship Type 1& ABBC (-3.88\%)& ABBA (+3.43\%)& ABAB (+1.36\%)\\
    Ship Type 2& AAAA (+2.50\%)& ABBB (-2.04\%)& AAAB (-2.02\%)\\
    Ship Type 3& ABBB (+2.67\%)& AAAB (-2.64\%)& AAAA (-2.63\%)\\
    Ship Type 4& ABAB (+1.86\%)& ABBB (-1.67\%)& AAAB (-1.64\%)\\
    Ship Type 5& AAAB (2.16\%)& ABBB (-2.08\%)& AAAA (-2.05\%)\\ 
    MSNBC & AAAB (-2.38\%)& AAAA (+2.34\%)& ABBA (+2.14\%)\\  
    Bike (L.A.) & ABCD (-0.46\%) & ABCA (+0.34\%)& ABBA (+0.15\%)\\
    Bike (Lon) & ABCA (+0.23\%)& AABC (-0.15\%)&AAAB (-0.13\%)\\ 
    US Flights & ABAB (-19.1\%)& ABCD (+9.32\%)& ABCA (+8.36\%)\\
    Taxi& ABBB (-1.95\%)& AAAA (+1.86\%)&AAAB (-1.77\%)\\
    \hline
\end{tabular}
\caption{Table with the worst represented $3$-hop motifs ranked in order of magnitude, calculated using the percentage observed counts minus the percentage expected counts. Positive percentages represent more observations in the data.}
\label{tab:worst-motif}
\end{table}
Our results for the worst motifs are shown in Table~\ref{tab:worst-motif}. Interestingly, we see that the $ABBB$ and $AAAB$ motif are consistently badly represent across 4 out of the 5 ship types, which suggests a common behaviour across these 4 ship types that result in a similar sequential correlation. Notably, $ABBB$ and $AAAB$ are motifs involving the shorter $AAA$ motif, which further implies that the correlation (and therefore the implicit mechanistic behaviour underlying the correlation) mainly stems from the transitions into and out of the $AAA$ motif. In order words, tendencies for ship to self-loop might involve even higher order correlations, or other more complex behaviours.
For both Bike datasets, the worst performing motifs still only show small deviations, which further suggests that there are little to no higher order correlations in the movement of these bikes. For the Flight dataset, the $ABAB$ motif is highly under-represented in the real data compared to simulations. This matches our expectations as we expect a high density of $ABA$ motifs due to the predominance of return flights, which results in an overestimation of $ABA$ chaining into $ABA$ in the $2^{nd}$ order model.

\subsubsection{Statistical Tests}
While the absolute and percentage counts can provide an intuitive glimpse into the expected versus observed frequencies of motif patterns in the various systems, it does not provide a statistical handle on whether these differences are expected under random statistical noise. Due to the sequential nature of the motifs, they are not independent and identically distributed (I.I.D) variables; the previous motif affects what the next motif could be. However, individual paths themselves are I.I.D. Therefore, by taking a sample of many paths, we can obtain a good estimate of the expected counts and variance of these motifs. This allows us to use the studentized residuals $R_M$ as defined in equation~\ref{eq:residuals}.


Assuming that the expected counts of each motif are approximately normally distributed, good agreement between observations and expectations occurs when $R_M \lesssim 1$. Non-normality in the motif count distributions will generally result in a larger calculated value of $R_M$ due to fat-tailed distributions. 

\begin{table}[h]
    \centering
    \arraycolsep=5pt\def\arraystretch{1.2}
    \begin{tabular}{|c|c|c|c|c|c|}
        \hline
         & & \multicolumn{4}{c|}{Studentized Residual $R_M$} \\
        \cline{3-6}
         & & \multicolumn{2}{c|}{2-hop motifs} & \multicolumn{2}{c|}{3-hop motifs}\\
         \cline{3-6} 
    & Order & Mean& Max & Mean& Max\\
        \hline
             Ship Type 1& 2  & 0.1&0.1& 1.0&\textcolor{red}{3.0}\\
        & 3  & 0.1&0.1& 0.1&0.1\\
        \hline
     Ship Type 2& 2  & 0.1&0.1&0.3&
    1.5\\
     & 3  & 0.1&0.1&0.1&
    0.1\\
    \hline
     Ship Type 3& 2  & 0.1&0.1&1.7&\textcolor{red}{6.6}\\
     & 3  & 0.1&0.1&0.1&0.1\\
     \hline
     Ship Type 4& 2  & 0.1&0.1&1.1&\textcolor{red}{4.2}\\
     & 3  & 0.1&0.1&0.1&0.1\\
     \hline
     Ship Type 5& 2  & 0.1&0.1&0.8&\textcolor{red}{2.8}\\
     & 3  & 0.1&0.1&0.1&0.1\\
     \hline
      MSNBC& 2  & 0.6&0.7&\textcolor{red}{14.2}&\textcolor{red}{95.8}\\
     & 3  & 0.5&1.3&0.6&1.5\\ 
     \hline
     Bike (L.A.)& 2  & 0.1&0.1&0.5&1.9\\
     & 3  & 0.1&0.2&0.2&0.4\\
    \hline
     Bike (Lon) & 2  & 0.2&0.3& 0.3&
    1.0\\
     & 3  & 0.2&0.5& 0.2&
    0.6\\
    \hline
     US Flights& 2  & \textcolor{red}{2.3}&\textcolor{red}{5.0}& \textcolor{red}{62.6}&
    \textcolor{red}{368.8}\\
     & 3  & 1.4&\textcolor{red}{2.6}& \textcolor{red}{3.9}&
    \textcolor{red}{7.9}\\
    \hline
     Taxi & 2  & 0.2&0.5& \textcolor{red}{6.1}&
    \textcolor{red}{16.5}\\
     & 3  & 0.4&0.5& 0.4&
    0.7\\
    \hline

\end{tabular}
\caption{Table showing the mean and max studentized residual $R_M$ of $2$ and $3$-hop motifs over order $2$ and $3$ higher-order networks, across all datasets. Numbers shaded in red represent residuals that are beyond an acceptable range of statistical uncertainty. $R_M \lesssim  1$ are seen as good fits between simulation and observation.}
\label{tab:residuals}
\end{table}

Our results in table~\ref{tab:residuals} show some interesting features. Firstly, for the ship type and bike datasets, the average $3$-hop motifs are still well represented at order $2$, which suggests that there are little to no higher order correlations beyond length 2, apart from 4 of the ship types which display at least one $3$-hop motif that has a statistically significant difference. The MSNBC and taxi datasets show higher order correlations, evident by the fact that the $3$-hop motifs are not well modelled by a $2^{nd}$ order network. The US flights residuals yield some unexpected results, where the $H$-hop motif at order $H$ still yields high values of $R_M$. This is likely due to the non-normality of the expected motif count distributions, where the variance does not capture the variability well. 


\section{Conclusions}
\label{Conclusion}
In this paper we have set out to investigate in detail what a model of optimal order entails when its performance is tested by comparing features of simulated and real data. This is a multifaceted task, as features of paths can be considered from micro- to macroscale, from frequency of motifs to path lengths and their distribution. We believe that this is an important stress test, since often the identification of the right model and the inference of its parameters is followed by a prediction stage, where the next move/state is forecasted. Our analysis has also revealed the need to find appropriate metrics to measure the quality of fit between simulated and real data. Selecting these metrics required careful consideration of what should be the appropriate choices, especially where no asymptotic results for the distributions of interest exist. 

More precisely, we were able to investigate the performance of higher-order networks at different orders, corresponding to the sequential correlation length, with the additional inclusion of an initial and final node. We then evaluated the dynamics predicted by the model via a random walk process on the higher-order network, measuring the sequence lengths and motif distribution compared to observations. Our results show that the model's ability to reproduce the sequence length distribution observed in real-world data is inconsistent, with certain datasets failing to converge at sufficiently low orders. Other datasets that do converge generally still require a higher order than would be expected from other information criteria, such as AIC and BIC. Meanwhile, the motif distributions measuring $H$-hop motif frequencies compared to the $(H-1)^{th}$ order predictions reveal that not all motifs are equally misrepresented. Certain motifs contribute heavily to the overall deviation, and therefore result in misrepresentation of the data by the model. Taken together, these two results show that the higher transition probabilities of order $k$, which assume dependence on the previous $k$ states, can represent some systems or some parts of a system well, but may struggle in other areas. Optimal orders as estimated by information criteria do not always guarantee a good representation of the system, and so a more careful evaluation of which measurements are important is required. 

During our analysis we also identified areas where this work could be extended. A significant but unaddressed issue in this paper is the time dimension in which these paths sit. Effectively, our work considers only the order of elements in a sequence and loses information about the time spent in one location or the time taken to reach another. It also treats the paths as happening independently, both of each other, and of the agents performing them. An additional feature worth considering in future analysis is other attributes of path hops: one feature that is relevant to  taxis/ships is which  path hops are empty, and which ones are carrying passengers/cargo,  respectively. In both of the latter examples, agents are motivated to minimise the time they spend moving without passengers/cargo, so it would be reasonable to assume that their next movements depend not just on where they were for the last $1, 2, \dots , k$ steps but whether each of those movements were empty or not. Future work should therefore consider including additional temporal characteristics and hop attributes in the models developed.

As part of this paper's contribution, we developed a Python package which can be installed from [GitHub repository will be made public upon first revision/acceptance]. With this package, users can efficiently extract different characteristics of their path data, including sequential motifs of arbitrary length, construct multi-order network models of a given maximal order $k$, and generate simulated paths from these models. Also provided in this repository are scripts for reproducing the experiments in this paper (apart from for the proprietary shipping data).


\section*{Acknowledgements}

Kevin Teo acknowledges the PhD studentship support from Northeastern University London. The authors would like to thank Dr Nicos Georgiou and Professor Ingo Scholtes for their useful discussions.
\newpage
\bibliographystyle{comnet}
\bibliography{references}

\begin{thebibliography}{00}

\bibitem{aic}
Akaike, H. (1998)  Information theory and an extension of the maximum likelihood principle. In {\em Selected papers of hirotugu akaike}, pages 199--213. Springer.

\bibitem{benson2021higher}
Benson, A.~R., Gleich, D.~F. {\&} Higham, D.~J. (2021)  Higher-order network analysis takes off, fueled by classical ideas and new data. {\em arXiv preprint arXiv:2103.05031}.

\bibitem{bick2023higher}
Bick, C., Gross, E., Harrington, H.~A. {\&} Schaub, M.~T. (2023)  What are higher-order networks?. {\em SIAM Review}, \textbf{65}(3), 686--731.

\bibitem{buhlmann1999variable}
B{\"u}hlmann, P. {\&} Wyner, A.~J. (1999)  Variable length Markov chains. {\em The Annals of Statistics}, \textbf{27}(2), 480--513.

\bibitem{bts2023odsurvey}
{Bureau of Transportation Statistics} (1993)  2023 Q1 Origin and Destination Survey. \url{https://www.transtats.bts.gov/DL_SelectFields.aspx?gnoyr_VQ=FLM&QO_fu146_anzr=b4vtv0%20n0q%20Qr56v0n6v10%20f748rB}.
Accessed: 2024-04-29.

\bibitem{deshpande2004selective}
Deshpande, M. {\&} Karypis, G. (2004)  Selective markov models for predicting web page accesses. {\em ACM transactions on internet technology (TOIT)}, \textbf{4}(2), 163--184.

\bibitem{ding2018modelselection}
Ding, J., Tarokh, V. {\&} Yang, Y. (2018)  Model selection techniques: An overview. {\em IEEE Signal Processing Magazine}, \textbf{35}(6), 16--34.

\bibitem{edler2017mapping}
Edler, D., Bohlin, L. {\&} Rosvall, M. (2017)  Mapping higher-order network flows in memory and multilayer networks with infomap. {\em Algorithms}, \textbf{10}(4), 112.

\bibitem{fieberg2024statistics}
Fieberg, J. (2024)  Statistics for Ecologists: A Frequentist and Bayesian Treatment of Modern Regression Models.. .

\bibitem{fournier2016spmf}
Fournier-Viger, P., Lin, J. C.-W., Gomariz, A., Gueniche, T., Soltani, A., Deng, Z. {\&} Lam, H.~T. (2016)  The SPMF open-source data mining library version 2. In {\em Machine Learning and Knowledge Discovery in Databases: European Conference, ECML PKDD 2016, Riva del Garda, Italy, September 19-23, 2016, Proceedings, Part III 16}, pages 36--40. Springer.

\bibitem{gote2020predicting}
Gote, C., Casiraghi, G., Schweitzer, F. {\&} Scholtes, I. (2020)  Predicting sequences of traversed nodes in graphs using network models with multiple higher orders. {\em arXiv preprint arXiv:2007.06662}.

\bibitem{gote2023predicting}
Gote, C., Casiraghi, G., Schweitzer, F. {\&} Scholtes, I. (2023)  Predicting variable-length paths in networked systems using multi-order generative models. {\em Applied Network Science}, \textbf{8}(1), 68.

\bibitem{grimmett2014probability}
Grimmett, G. {\&} Welsh, D.~J. (2014) {\em Probability: an introduction}.
Oxford University Press.

\bibitem{grinstead2012introduction}
Grinstead, C.~M. {\&} Snell, J.~L. (2012) {\em Introduction to probability}.
American Mathematical Soc.

\bibitem{kovanen2011temporal}
Kovanen, L., Karsai, M., Kaski, K., Kert{\'e}sz, J. {\&} Saram{\"a}ki, J. (2011)  Temporal motifs in time-dependent networks. {\em Journal of Statistical Mechanics: Theory and Experiment}, \textbf{2011}(11), P11005.

\bibitem{kullback1951information}
Kullback, S. {\&} Leibler, R.~A. (1951)  On information and sufficiency. {\em The annals of mathematical statistics}, \textbf{22}(1), 79--86.

\bibitem{lambiotte2019networks}
Lambiotte, R., Rosvall, M. {\&} Scholtes, I. (2019)  From networks to optimal higher-order models of complex systems. {\em Nature physics}, \textbf{15}(4), 313--320.

\bibitem{larock2022sequential}
LaRock, T., Scholtes, I. {\&} Eliassi-Rad, T. (2022)  Sequential motifs in observed walks. {\em Journal of Complex Networks}, \textbf{10}(5), cnac036.

\bibitem{lewis2017simulation}
Lewis, P. {\&} McKenzie, E. (2017) {\em Simulation Methodology for Statisticians, Operations Analysts, and Engineers (1988)}.
Chapman and Hall/CRC.

\bibitem{labikesharemetro}
{Metro Bike Share}.
\url{https://bikeshare.metro.net/about/data/}.

\bibitem{milo2002network}
Milo, R., Shen-Orr, S., Itzkovitz, S., Kashtan, N., Chklovskii, D. {\&} Alon, U. (2002)  Network motifs: simple building blocks of complex networks. {\em Science}, \textbf{298}(5594), 824--827.

\bibitem{taxiECMLPKDD2015}
Moreira-Matias, L., Ferreira, M., Mendes-Moreira, J., L, L. {\&} J, J. (2015)  {Taxi Service Trajectory - Prediction Challenge, ECML PKDD 2015}. UCI Machine Learning Repository.
{DOI}: https://doi.org/10.24432/C55W25.

\bibitem{bic}
Neath, A.~A. {\&} Cavanaugh, J.~E. (2012)  The Bayesian information criterion: background, derivation, and applications. {\em Wiley Interdisciplinary Reviews: Computational Statistics}, \textbf{4}(2), 199--203.

\bibitem{oeis}
{OEIS Foundation Inc.} (2024)  The {O}n-{L}ine {E}ncyclopedia of {I}nteger {S}equences. Published electronically at \url{https://oeis.org/A000110}.

\bibitem{paranjape2017motifs}
Paranjape, A., Benson, A.~R. {\&} Leskovec, J. (2017)  Motifs in temporal networks. In {\em Proceedings of the tenth ACM international conference on web search and data mining}, pages 601--610.

\bibitem{peixoto2017modelling}
Peixoto, T.~P. {\&} Rosvall, M. (2017)  Modelling sequences and temporal networks with dynamic community structures. {\em Nature communications}, \textbf{8}(1), 582.

\bibitem{petrovic2022learning}
Petrovic, L.~V. {\&} Scholtes, I. (2022)  Learning the Markov order of paths in graphs. In {\em Proceedings of the ACM web conference 2022}, pages 1559--1569.

\bibitem{tfl2023bikejourney}
{Powered by TfL Open Data} (1993) .
\url{https://cycling.data.tfl.gov.uk}.
`Contains OS data © Crown copyright and database rights 2016' and Geomni UK Map data © and database rights [2019], Accessed: 2024-05-20.

\bibitem{rosvall2014memory}
Rosvall, M., Esquivel, A.~V., Lancichinetti, A., West, J.~D. {\&} Lambiotte, R. (2014)  Memory in network flows and its effects on spreading dynamics and community detection. {\em Nature communications}, \textbf{5}(1), 4630.

\bibitem{saebi2020efficient}
Saebi, M., Xu, J., Kaplan, L.~M., Ribeiro, B. {\&} Chawla, N.~V. (2020)  Efficient modeling of higher-order dependencies in networks: from algorithm to application for anomaly detection. {\em EPJ Data Science}, \textbf{9}(1), 15.

\bibitem{scholtes2017network}
Scholtes, I. (2017)  When is a network a network? Multi-order graphical model selection in pathways and temporal networks. In {\em Proceedings of the 23rd ACM SIGKDD international conference on knowledge discovery and data mining}, pages 1037--1046.

\bibitem{schwarz1978estimating}
Schwarz, G. (1978)  Estimating the dimension of a model. {\em The annals of statistics}, pages 461--464.

\bibitem{tonon2023caspita}
Tonon, A. {\&} Vandin, F. (2023)  caSPiTa: mining statistically significant paths in time series data from an unknown network. {\em Knowledge and Information Systems}, \textbf{65}(6), 2347--2374.

\bibitem{trench2012lagrange}
Trench, W.~F. (2012)  The Method of Lagrange Multipliers. {\em Research Gate, Book}.

\bibitem{wilks1938large}
Wilks, S.~S. (1938)  The large-sample distribution of the likelihood ratio for testing composite hypotheses. {\em The annals of mathematical statistics}, \textbf{9}(1), 60--62.

\bibitem{chawla2016representing}
Xu, J., Wickramarathne, T.~L. {\&} Chawla, N.~V. (2016)  Representing higher-order dependencies in networks. {\em Science advances}, \textbf{2}(5), e1600028.

\end{thebibliography}

\end{document}